\documentclass[longauth]{aa}
\usepackage{hyperref}

\usepackage{natbib}
\bibpunct{(}{)}{;}{a}{,}{,}
\usepackage{amsmath}
\usepackage{graphicx}
\usepackage{txfonts}

\usepackage[usenames,dvipsnames,svgnames,table]{xcolor}

\usepackage[super]{nth}
\usepackage{ulem}

\usepackage{subfig}
\usepackage{multirow}

\newcommand{\eg}{{\it e.g.}~}

\newcommand{\Sinv}{{\bf \mathsf{ S}}^{-1}}
\newcommand{\Sig}{{\bf \mathsf{ S}}}

\newcommand{\B}{\mathcal{B}}
\newcommand{\K}{\mathcal{K}}
\newcommand{\E}{\mathcal{E}}

\newcommand{\D}{\mathcal{D}}

\newcommand{\bdelta}{{\bf \delta}}

\newcommand{\bN}{{\bf N}}
\newcommand{\Nbar}{\bar{N}}

\newcommand{\LCDM}{$\Lambda$CDM}
\newcommand{\hmpc}{$h^{-1}$Mpc}
\newcommand{\mpch}{$h$Mpc$^{-1}$}

\newcommand{\according}{$\leftarrow$}

\begin{document}
\title{The VIMOS Public Extragalactic Redshift Survey}
\subtitle{Reconstruction of the redshift-space galaxy density field}
\titlerunning{VIPERS reconstruction of the redshift-space galaxy density field}

\author{B.~R.~Granett\inst{\ref{2}}
\and E.~Branchini\inst{\ref{10},\ref{28},\ref{29}}
\and L.~Guzzo\inst{\ref{2},\ref{27}}
\and U.~Abbas\inst{\ref{5}}
\and C.~Adami\inst{\ref{4}}
\and S.~Arnouts\inst{\ref{6}}
\and J.~Bel\inst{\ref{2}}
\and M.~Bolzonella\inst{\ref{9}}
\and D.~Bottini\inst{\ref{3}}
\and A.~Cappi\inst{\ref{9},\ref{21}}
\and J.~Coupon\inst{\ref{12}}
\and O.~Cucciati\inst{\ref{17},\ref{9}}           
\and I.~Davidzon\inst{\ref{9},\ref{17}}
\and G.~De Lucia\inst{\ref{13}}
\and S.~de la Torre\inst{\ref{4}}
\and A.~Fritz\inst{\ref{3}}
\and P.~Franzetti\inst{\ref{3}}
\and M.~Fumana\inst{\ref{3}}
\and B.~Garilli\inst{\ref{3},\ref{4}}
\and O.~Ilbert\inst{\ref{4}}
\and A.~Iovino\inst{\ref{2}}
\and J.~Krywult\inst{\ref{15}}
\and V.~Le Brun\inst{\ref{4}}
\and O.~Le F\`evre\inst{\ref{4}}
\and D.~Maccagni\inst{\ref{3}}
\and K.~Ma{\l}ek\inst{\ref{23}}
\and F.~Marulli\inst{\ref{17},\ref{18},\ref{9}}
\and H.~J.~McCracken\inst{\ref{19}}
\and M.~Polletta\inst{\ref{3}}
\and A.~Pollo\inst{\ref{22},\ref{23}}
\and M.~Scodeggio\inst{\ref{3}}
\and L.~A.~M.~Tasca\inst{\ref{4}}
\and R.~Tojeiro\inst{\ref{30}}
\and D.~Vergani\inst{\ref{25},\ref{9}}
\and A.~Zanichelli\inst{\ref{26}}
\and A.~Burden\inst{\ref{11}}
\and C.~Di Porto\inst{\ref{9}}
\and A.~Marchetti\inst{\ref{3}}
\and C.~Marinoni\inst{\ref{7}}
\and Y.~Mellier\inst{\ref{19}}
\and T.~Moutard\inst{4}
\and L.~Moscardini\inst{\ref{17},\ref{18},\ref{9}}
\and R.~C.~Nichol\inst{\ref{11}}
\and J.~A.~Peacock\inst{\ref{14}}
\and W.~J.~Percival\inst{\ref{11}}
\and G.~Zamorani\inst{\ref{9}}
}
\offprints{Benjamin R. Granett \\ \email{ben.granett@brera.inaf.it}}
\institute{
INAF - Osservatorio Astronomico di Brera, Via Brera 28, 20122 Milano, via E. Bianchi 46, 23807 Merate, Italy \label{2}
\and Dipartimento di Matematica e Fisica, Universit\`{a} degli Studi Roma Tre, via della Vasca Navale 84, 00146 Roma, Italy \label{10}
\and INFN, Sezione di Roma Tre, via della Vasca Navale 84, I-00146 Roma, Italy \label{28}
\and INAF - Osservatorio Astronomico di Roma, via Frascati 33, I-00040 Monte Porzio Catone (RM), Italy \label{29}
\and Dipartimento di Fisica, Universit\`a di Milano-Bicocca, P.zza della Scienza 3, I-20126 Milano, Italy \label{27}
\and INAF - Osservatorio Astronomico di Torino, 10025 Pino Torinese, Italy \label{5}
\and Aix Marseille Universit\'e, CNRS, LAM (Laboratoire d'Astrophysique de Marseille) UMR 7326, 13388, Marseille, France  \label{4}
\and Canada-France-Hawaii Telescope, 65--1238 Mamalahoa Highway, Kamuela, HI 96743, USA \label{6}
\and INAF - Osservatorio Astronomico di Bologna, via Ranzani 1, I-40127, Bologna, Italy \label{9}
\and INAF - Istituto di Astrofisica Spaziale e Fisica Cosmica Milano, via Bassini 15, 20133 Milano, Italy\label{3}
\and Laboratoire Lagrange, UMR7293, Universit\'e de Nice Sophia Antipolis, CNRS, Observatoire de la C\^ote d’Azur, 06300 Nice, France \label{21}
\and Astronomical Observatory of the University of Geneva, ch. d'Ecogia 16, CH-1290 Versoix, Switzerland \label{12}
\and Dipartimento di Fisica e Astronomia - Alma Mater Studiorum Universit\`{a} di Bologna, viale Berti Pichat 6/2, I-40127 Bologna, Italy \label{17}
\and INAF - Osservatorio Astronomico di Trieste, via G. B. Tiepolo 11, 34143 Trieste, Italy \label{13}
\and Institute of Physics, Jan Kochanowski University, ul. Swietokrzyska 15, 25-406 Kielce, Poland \label{15}
\and National Centre for Nuclear Research, ul. Hoza 69, 00-681 Warszawa, Poland \label{23}
\and INFN, Sezione di Bologna, viale Berti Pichat 6/2, I-40127 Bologna, Italy \label{18}
\and Institute d'Astrophysique de Paris, UMR7095 CNRS, Universit\'{e} Pierre et Marie Curie, 98 bis Boulevard Arago, 75014 Paris, France \label{19}
\and Astronomical Observatory of the Jagiellonian University, Orla 171, 30-001 Cracow, Poland \label{22}
\and School of Physics and Astronomy, University of St Andrews, North Haugh,St Andrews KY16 9SS, UK \label{30}
\and INAF - Istituto di Astrofisica Spaziale e Fisica Cosmica Bologna, via Gobetti 101, I-40129 Bologna, Italy \label{25}
\and INAF - Istituto di Radioastronomia, via Gobetti 101, I-40129, Bologna, Italy \label{26}
\and Institute of Cosmology and Gravitation, Dennis Sciama Building, University of Portsmouth, Burnaby Road, Portsmouth, PO1 3FX \label{11}
\and Centre de Physique Th\'eorique, UMR 6207 CNRS-Universit\'e de Provence, Case 907, F-13288 Marseille, France \label{7}
\and SUPA, Institute for Astronomy, University of Edinburgh, Royal Observatory, Blackford Hill, Edinburgh EH9 3HJ, UK \label{14}}


\abstract{}
{Using the VIMOS Public Extragalactic Redshift Survey (VIPERS) we aim to jointly estimate the key parameters that describe the galaxy density field and its spatial correlations in redshift space.}
{We use the Bayesian formalism to jointly reconstruct the redshift-space galaxy density field, power spectrum,  galaxy bias and galaxy luminosity function given the observations and survey selection function.  The high-dimensional posterior distribution is explored using the Wiener filter within a Gibbs sampler.
 We validate the analysis using simulated catalogues and apply it to VIPERS data taking into consideration the inhomogeneous selection function.}
{We present joint constraints on the anisotropic power spectrum as well as the bias and number density of red and blue galaxy classes in luminosity and redshift bins as well as the measurement covariances of these quantities.  We find that the inferred galaxy bias and number density parameters are strongly correlated although these are only weakly correlated with the galaxy power spectrum.  The power spectrum and redshift-space distortion parameters are in agreement with previous VIPERS results with the value of the growth rate $f\sigma_8=0.38$ with 18\% uncertainty at redshift 0.7.}
{}
\maketitle

\section{Introduction}

The distribution of galaxies on large scales provides a fundamental test of the cosmological model.  In the standard picture, galaxies trace an underlying matter density field and the statistical properties of this field such as its power spectrum and higher order moments are given by the theory \citep{Peebles}.  This clear view is confounded, however, by the sparse distributions of luminous galaxies mapped by surveys \citep{Lahav04}.  Galaxies are complex systems; they are biased tracers of the non-linear matter field and their clustering strength depends on their properties and formation histories \citep{davis76,blanton05,kaiser84,bardeen86,mo96}.  Furthermore, their redshift gives a distorted view of distance, affected by coherent and random velocities \citep{kaiser87}.

Upcoming galaxy surveys will be sensitive to subtleties in these trends  requiring increasingly sophisticated modelling and numerical simulations to interpret the galaxy distribution in detail \citep{des,lsst,desi,euclid}.
At the same time, these surveys will be  sufficiently large to be limited by minute selection effects that systematically and significantly alter the observed distribution of galaxies.  Instrumental and observational artefacts can  masquerade as genuine astrophysical effects and vice-versa.  Thus the analyses will  need to track  a large number of instrumental and astrophysical  parameters and be able to characterise the covariances between them.  Reliable error estimation will require incorporating the set of both systematic and random uncertainties.  The stakes are high as experiments promise highly precise  constraints on the nature of gravity, dark energy and dark matter \citep{euclidcosmo}.

Together, the physical and instrumental models compose the total data model.  Given the large number of parameters, the Bayesian approach is often preferred over the frequentist one to set joint constraints on the relevant physical quantities \citep{trotta08}.   At the heart of this approach is Bayes theorem which dictates a recipe for translating a set of observations into constraints on model parameters.  Of fundamental importance is the incorporation of any prior knowledge of these parameters.  This framework  provides a natural means to jointly constrain  physical parameters of interest while marginalising over a set of nuisance parameters.  A paradigmatic example is the analysis of cosmic microwave background data \citep{Wandelt04-CMB} as demonstrated through the ESA Planck mission results \citep{planck}.  

The Wiener filter is the first example of the application of Bayesian reconstruction techniques to galaxy surveys.   The Wiener solution corresponds to the maximum {\it a posteriori} solution given a Gaussian likelihood and prior.  In general, for a signal contaminated by noise, the Wiener filter gives a reconstruction of the true signal with the minimum residual variance \citep{Rybicki04}. This is true also for non-Gaussian signal and noise sources, and for this reason, since the galaxy field is not Gaussian (it is thought to tend toward Gaussianity on very large scales), the Wiener filter has seen significant use in reconstructing the density field from galaxy surveys \citep[\eg][]{Lahav94} and in particular to predict large-scale structures behind the Galactic plane \citep{Zaroubi95}.

Wiener filtering is comparable to other adaptive density reconstruction techniques such as  Delaunay tessellations \citep{Schaap00}, although Wiener filtering offers the advantage of naturally accounting for a complex survey selection function with inhomogeneous sampling.      Examples of applications of Wiener filtering to galaxy surveys include the  Two-degree Field Galaxy Redshift Survey (2dFGRS) in which the Wiener filter was used to identify galaxy clusters and voids \citep{Erdogdu04}.  \citet{Kitaura09-SDSS} present a Wiener density field reconstruction of the Sloan Digital Sky Survey (SDSS) main sample.  Applied to the VIMOS Extragalatic Redshift Survey (VIPERS; Guzzo et al. 2014), the Wiener filter can naturally account for  inhomogeneous sampling and survey gaps.  In a comparison study of different density field estimators for VIPERS  \citet{Cucciati14} find that the Wiener filter performs well although it over-smooths the field in low density environments affecting cell-count statistics.

\begin{figure}
\includegraphics[scale=0.9]{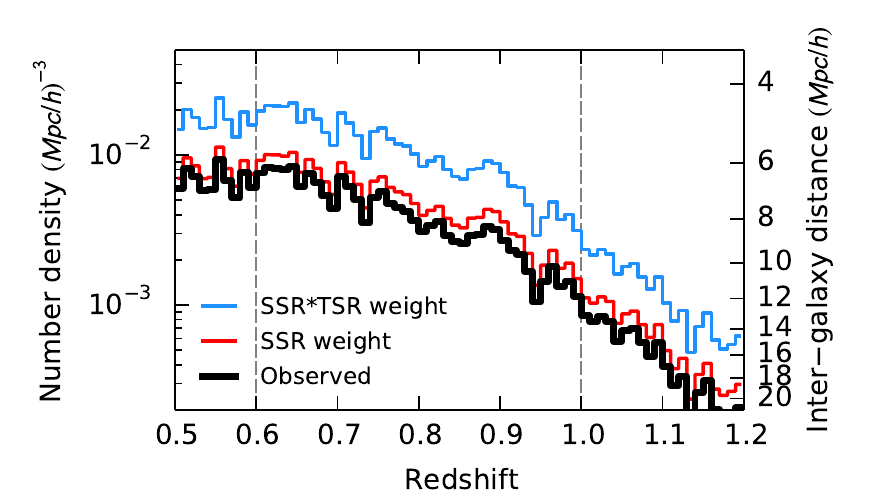}
\caption{The VIPERS galaxy number density of the v5 internal release sample. The curves show the effect of the completeness corrections including the spectroscopic success rate (SSR) and target sampling rate (TSR).  Our analysis uses the redshift range 0.6-1.0 (vertical lines). \label{fig:nz}}

\end{figure}

More physically motivated probability distribution functions have been developed to improve over the Wiener filter and obtain unbiased density field reconstructions.  \citet{Kitaura10-LN} demonstrate in a comparison study that the use of 
a Poisson sampling model for the galaxy counts with a log-normal prior on the density field allows better estimation of the lowest and highest density extremes on small scales.  The generalisation of the model calls for a fully non-linear solver \citep{Jasche10-HamSamp}. The Poisson log-normal model was used to reconstruct the density field probed by the SDSS sample \citep{Jasche10-SDSSLNP}.

The Gaussian likelihood has also been used to construct maximum  {\it a posteriori}   estimators for the galaxy power spectrum \citep{Efstathiou01M,Tegmark02,Pope04,Granett12}.  Also for the galaxy luminosity function estimates, maximum likelihood techniques have enjoyed significant use \citep{Ilbert05,Blanton03,Efstathiou88}.

Gaussian likelihood methods have only recently been developed to jointly infer the density field, power spectrum and luminosity function from galaxy surveys \citep{Kitaura08,Ensslin09}.  The first application to the Sloan Digital Sky Survey was demonstrated by \citet{Jasche10-SDSSBayes} who utilise a Gaussian likelihood and prior to jointly estimate the underlying galaxy field and power spectrum.  This work was generalised to simultaneously  estimate the linear galaxy bias and luminosity function \citep{Jasche13}.  \citet{Ata15} further model a scale-dependent and stochastic galaxy bias using the Log-normal Poisson model.  The methodology has also been developed for photometric redshift surveys \citep{Jasche12photo}.

The full description of the galaxy field requires consideration of the higher order moments and depends on the physics of structure formation.  Thus reconstruction methods have been developed that incorporate  physical models based on second-order perturbation theory \citep{Kitaura12-2MASS,Kitaura13-IC,Jasche13-IC,Jasche15-ICSDSS} or full N-body calculations \citep{Wang14}.  Reconstructions of the local Universe have been used in novel ways, including to estimate the bias in the Hubble constant due to cosmic flows \citep{Hess14}.

In this work we carry out a Bayesian analysis of the VIMOS Extragalactic Redshift Survey (VIPERS; Guzzo et al 2014).  Our goal is to jointly estimate the key statistics including the matter power spectrum, galaxy biasing function and galaxy luminosity function.
Our strategy is, given the observed number density of galaxies in the survey as a function of position $N(RA,Dec,z)$,  to compute the conditional probability distribution for the parameters, written schematically as: the matter over-density field $\delta$, galaxy mean number density $\bar{N}$, galaxy bias $b$ and the two-point correlation function $S$.   The conditional probability distribution or posterior may be decomposed using Bayes theorem:
\begin{equation}\label{eq:bayes}
p(\delta,\bar{N},b,S | N) \propto p(N|\delta,\bar{N},b,S)p(\delta,\bar{N},b,S).
\end{equation}
The first and second factors on the right-hand side are the data likelihood and the parameter prior.  We will account for observational systematics such as the survey selection function in the data model, but we will not propagate their uncertainties.  For VIPERS, the uncertainties in the selection function are subdominant compared with the statistical errors and so  the inclusion of the uncertainties will be reserved for future work.  We adopt the Gibbs sampling algorithm to sample from the posterior distribution \citep{bayesiancore}.  With this approach the complex joint probability distribution is broken up in a number of simpler, individual conditional distributions. Sampling these distributions allows us to build up a Markov chain that rapidly converges to the joint distribution.

 VIPERS has mapped the galaxy field to redshift 1 with unprecedented fidelity \citep{Guzzo14,Garilli14}.  So far, VIPERS data have been used to constrain the growth rate of structure through the shape of the redshift-space galaxy correlation function  \citep{Delatorre13}. The cosmological interpretation of the galaxy power spectrum monopole has been presented by Rota et al (in prep.).  The biasing function that links the galaxy and dark matter density has been estimated using the luminosity-dependent correlation function \citep{Marulli13} and the shape of the 1-point probability distribution function of galaxy counts in cells  \citep{Diporto14}.  Moreover VIPERS has tightened the constraints on the galaxy luminosity and stellar mass functions \citep{Davidzon13,Fritz14}.
These measurements firmly anchor models of galaxy formation at redshift 1.

Beyond the one and two point statistics of the galaxy field, galaxies are organised  into a cosmic web of knots, filaments and walls that surround large empty voids.   In VIPERS the higher order moments of the galaxy counts in cells distribution function have been measured (Cappi et al, 2015).
While ongoing efforts are being made to measure the morphologies of cosmic structures.    A catalogue of   voids has been constructed with VIPERS (Micheletti et al., 2014; Hawken et al., in prep).  More generally, the Minkowski functionals may be used to characterise the topology of large-scale structure  as a function of scale (Schimd, {\it in prep.}).  These measurements typically call for precise reconstructions of the density field corrected for observational systematics such as survey gaps and inhomogeneous sampling.

This work extends previous analyses by considering the joint distribution of galaxy luminosity, colour and clustering bias with the spatial power spectrum and density field.   We begin in Sec.~\ref{sec:vipers} with an overview of VIPERS and the parameterisation of the selection function.  The data model is described in Sec.~\ref{sec:datamodel}, and the method is outlined in \ref{sec:method}.    In Sec \ref{sec:results} we present the constraints from VIPERS data.

We assume the following fiducial cosmology $\Omega_m=0.27$, $\Omega_b=0.0469$, $\Omega_\Lambda=0.73$, $n_s=0.95$, $H_0=70$ km/s/Mpc, $\sigma_8=0.80$.  This coincides with the MultiDark simulation run \citep{Prada12} that was used to construct the mock VIPERS catalogues.   Magnitudes are in the AB system unless noted.  The absolute magnitudes used were computed under a flat cosmology with $\Omega_m=0.30$; however we transform all magnitudes to the $\Omega_m=0.27$ cosmology.

\begin{figure}
\hspace{-0.1in}\includegraphics[scale=0.9]{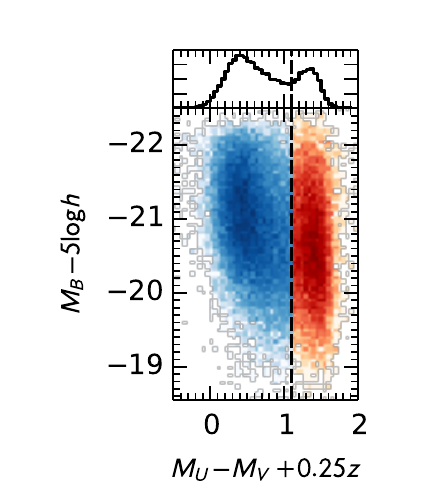}\hspace{-0.2in} \includegraphics[scale=0.9]{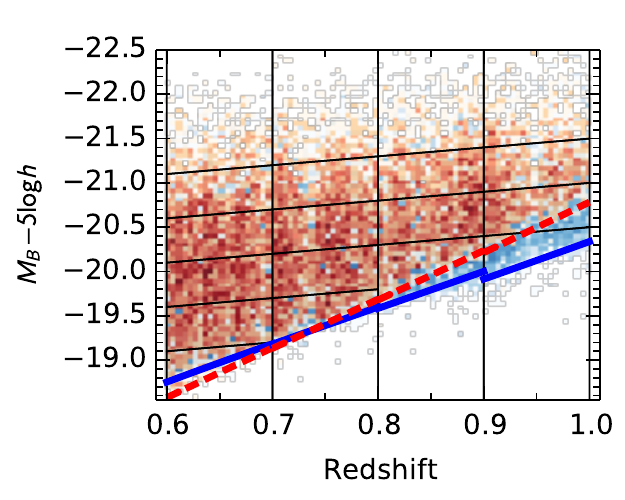}
\caption{We construct subsamples of galaxies in colour, absolute magnitude and redshift bins.  Left: we show the absolute magnitude-colour plane.  The histogram at top shows the distribution of colour.  We divide the sample into blue and red classes following \citet{Fritz14}  at $M_U-M_V + 0.25z=1.1$.  Right: the sample is further binned by redshift and absolute magnitude.  The luminosity bins account for the mean evolutionary trend.  The faintest luminosity bins are not volume limited and the thick blue and dashed red curves show the limiting magnitudes for blue and red galaxy classes.
\label{fig:samples}}
\end{figure}

 \begin{figure*}
\includegraphics{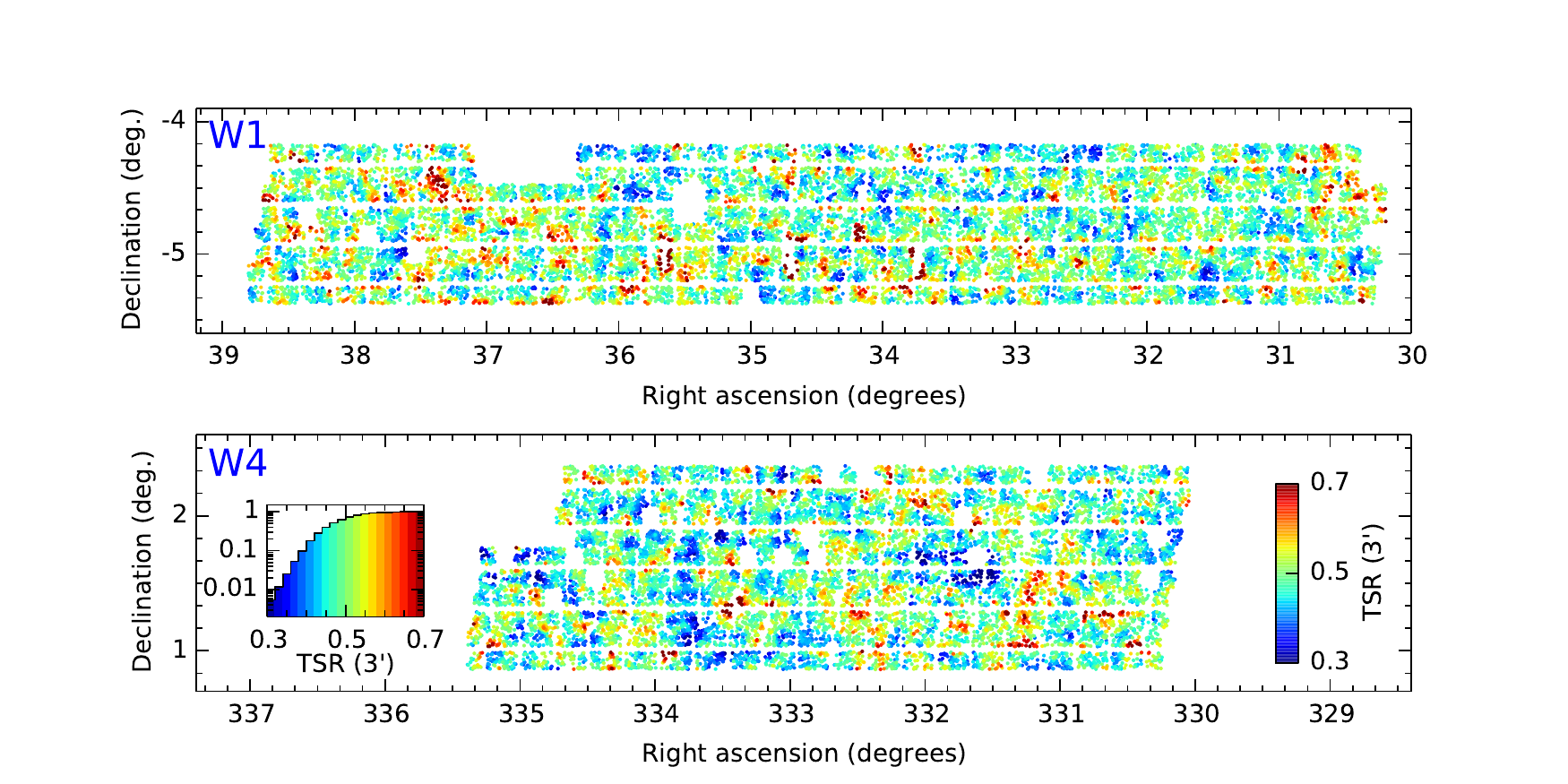}
\caption{The distribution of VIPERS targets on the sky is plotted for the two fields W1 and W4.    The points are coloured according to the target sampling rate (TSR) which is defined as the ratio of the number of targeted galaxies in a patch of sky over the total number available in the parent photometric catalogue.   In this work we estimate the TSR within a circular aperture with radius 3 arcmin.  The TSR depends on the projected density of targets on the sky.  TSR is higher in low-density fields with few potential targets while in high density fields slit positioning constraints severely limit the number of sources that may be targeted.  The inset histogram shows the cumulative fraction of targets with TSR below a given value.  The median TSR over the survey is 48\%. \label{fig:tsr}}
\end{figure*}

  \begin{figure*}
\includegraphics[height=2.5in]{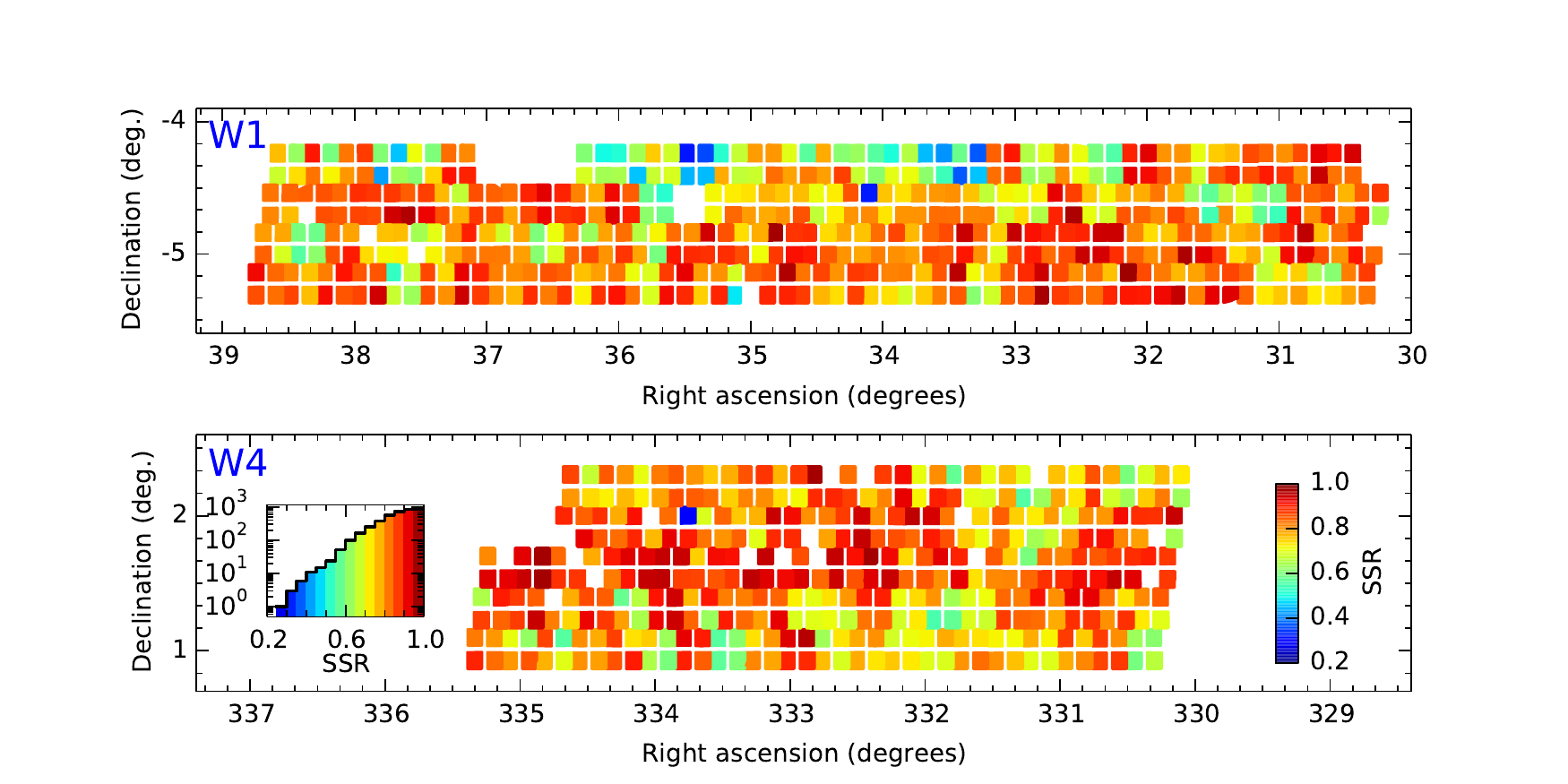}
\hspace{-.5in} \includegraphics[height=2.5in]{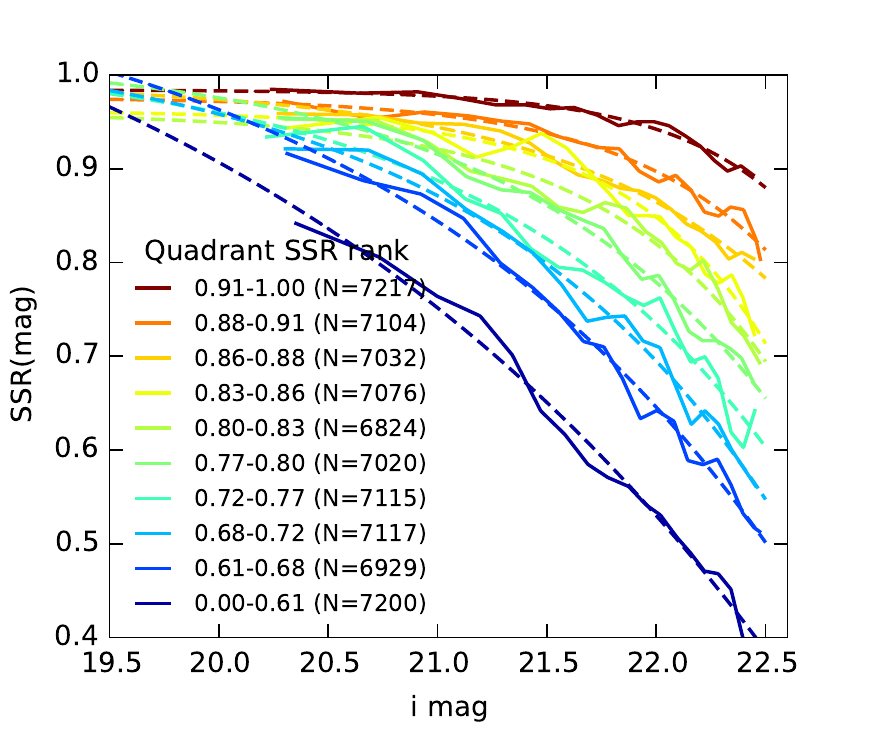}
\caption{The spectroscopic success rate (SSR) quantifies the fraction of targets for which the redshift could be measured with $>95\%$ confidence.  At left we show the mean SSR of each pointing divided by quadrant (7x8 arcmin).  The inset histogram gives the cumulative number of quadrants with SSR below the given value.  At right, we show the SSR as a function of $i_{AB}$ magnitude (solid curves) with an analytic fit (dashed curves).  The sample is divided based on the overall quality of the quadrants quantified by SSR.  The quadrants are ranked by mean SSR and the curves are computed for each decile.  The range of SSR and the number of sources in the bin are given in the figure legend.
\label{fig:ssr}}
\end{figure*}

\section{VIPERS}\label{sec:vipers}
The VIMOS Public Extragalactic Redshift Survey (VIPERS) is an ESO programme on VLT (European Souther Observatory - Very Large Telescope; \citealp{Guzzo14,Garilli14}).   The survey targets galaxies for medium resolution spectroscopy using  VIMOS (VIsible Multi-Object Spectrograph;  \citealp{vimos}) within two regions of the W1 and W4 fields of the CFHTLS-Wide Survey (Canada-France-Hawaii Telescope Legacy Wide;\citealp{cfht}).  Targets are chosen based upon colour selection to be in the redshift range 0.5-1.2.  The final expected sky coverage of VIPERS is $24 \, {\rm  deg}^2\!$.

For each galaxy, the $B$-band rest-frame magnitude was estimated following the Spectral Energy Distribution (SED) fitting method described in \citet{Davidzon13} and adopted to define volume  limited samples. The choice of $B$-band rest-frame is natural,  corresponding to the observed $I$-band at redshift $\sim 0.8$.  We derived K-corrections from the best-fitting SED templates using all available photometry including near-UV, optical, and near-infrared.
 
 \subsection{Sample selection}
 This analysis is based on the VIPERS v5 internal data release which represents 77\% of the final survey.
We select sources from the VIPERS catalogue in the redshift range $0.6-1.0$ with redshift confidence $>95\%$ (redshift flags 2,3,4,9).  The redshift distribution is shown in Fig. \ref{fig:nz}.
The sources have estimated rest frame Buser B-band magnitudes and  $U-V$ colours defined in the Johnson-Cousins-Kron system as described by \citet{Fritz14}.  The total number of sources used in the analysis is 36928.
We construct subsamples of galaxies in bins of redshift ($\Delta z=0.1$), luminosity  ($\Delta M_B=0.5$), and colour as illustrated in Fig.~\ref{fig:samples}.  We separate red and blue galaxy classes using the cut defined by \citet{Fritz14} at  $(M_U-M_V)_{\rm Vega} + 0.25 z = 1.1$ in the Vega system.   
In total there are 37 bins, 19 for blue  and 18 for red galaxy subsamples, including those that are not complete, see the discussion in Section 3.3.

 \subsection{Survey completeness}
 The VIPERS survey coverage is characterised by an angular mask \citep{Guzzo14}.  The mask is made up of a mosaic of VIMOS pointings, each consisting of four quadrants.  Regions around bright stars and of poor photometric quality in CFHTLS photometric catalogue have been removed.
 
Within an observed quadrant there are many factors including the intrinsic source properties, instrument response and observing conditions that determine the final selection function \citep{Garilli14}.  The fraction of sources out of the parent photometric sample that are targeted for spectroscopy 
is referred to as the target sampling rate (TSR).  Among the targeted sources, not all will give a reliable redshift measurement.  We refer to this fraction as the spectroscopic success rate (SSR).   The sampling rate is the product of the TSR and SSR:
$r = r_{TSR} r_{SSR}$.

The arrangement of slits in VIMOS is strongly constrained since the spectra cannot overlap on the imaging plane \citep{Bottini05}.  In VIPERS, the result is that the number  of targeted sources in a pointing is approximately constant, damping the galaxy clustering signal both on small and large scales \citep{Pollo05,Delatorre13}.

Since, as described in Section 3, in our data model we bin the galaxies onto a cubic grid, it is only necessary to estimate the TSR correction on the scale of the cubic cell.  For 5\hmpc\ cells, this corresponds to 10 arcmin at $z=0.7$, larger in fact, than the VIMOS quadrants.  The TSR is estimated on a fine grid as the fraction of targets out of the parent sample within a 3 arcmin circular aperture: $N_{target}/N_{parent}$.  The fine grid is then down-sampled to determine the average TSR in each grid cell.  The colours in Fig.~\ref{fig:tsr} indicate the TSR measurements as a function of angular position (at the positions of observed galaxies).

The spectroscopic success rate (SSR) is primarily correlated to the  conditions at the time of observation and so varies with pointing.   For a particular source the SSR depends on the apparent flux as well as the spectral features that are available to make the redshift measurement.  We find that the primary contribution comes from the apparent flux, and we quantify the mean SSR, defined as $N_{measured}/N_{target}$ in each quadrant, as a function of the $i$ band magnitude \citep{Guzzo14,Garilli14}.  The degradation is most severe in poor observing conditions, so we compute the SSR separately according to the quality of the quadrant.  We rank the quadrants based on mean SSR and compute separately $r_{SSR}(m_i)$ in each decile.  The SSR is fit with an analytic form: $r_{SSR}(m) = a (1-e^{c (m-b)})$.  In the right panel of Fig.~\ref{fig:ssr} we see that SSR depends strongly on the quadrant quality.  For the top 10\% of quadrants (shown by the red curve in Fig.~\ref{fig:ssr}), the SSR remains $>90\%$, but it drops quickly as quadrant quality falls.  What is important to note here is that the shape of the $r_{SSR}(i_{AB})$ curves changes as a function of quadrant quality.

The effect of the weights on the redshift distribution is shown in Fig. \ref{fig:nz}.

\subsection{Mock catalogues}
We use a set of simulated (mock) galaxy catalogues constructed to match the VIPERS observing strategy.
The catalogues are built on the MultiDark N-body simulation \citep{Prada12} using the Halo Occupation Distribution (HOD) technique.  Details of the construction  may be found in \citep{Delatorre13,Delatorre13b}.

Galaxies were added to the dark matter halos in the simulation according to a luminosity dependent HOD model.  The correlation function and number counts in luminosity bins were set to
match measurements made at $0.5<z<1.2$ in CFHTLS, VVDS and earlier releases of VIPERS.  Each mock galaxy is characterised by its angular coordinate, comoving distance, ``observed'' redshift including its errors and an absolute magnitude in the B band.  

We partition the mock catalogues into bins of redshift and luminosity, but not in colour, as we do for the VPERS data.  The step size in redshift and luminosity are $\Delta z=0.1$ and $\Delta M_B=0.2$ over the redshift range $0.6<z<1.0$.  We have 46 bins including those bins which are not complete due to the apparent flux limit.

The mock galaxies match the number density of the VIPERS observations, but do not include the slit placement constraints that we correct in the data with the TSR weights.  Nor do the catalogues simulate the spectroscopic sampling rate.

\begin{figure*}
 \hspace{-.2in}\includegraphics[scale=0.8]{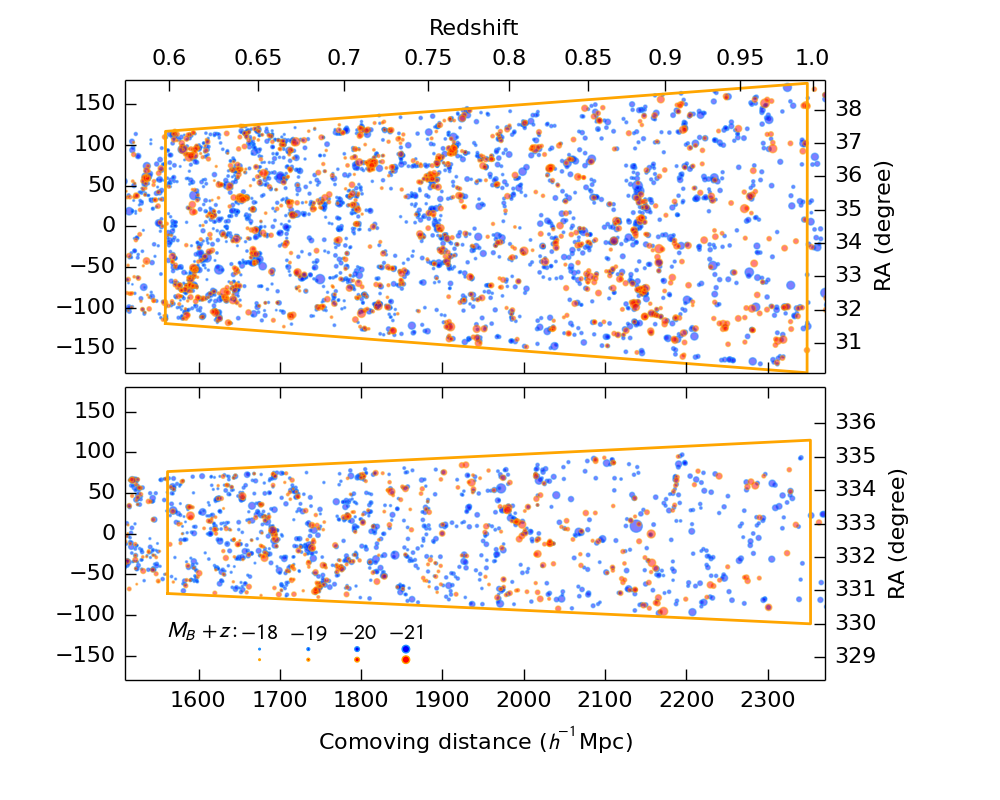}
 \hspace{-.6in} \includegraphics[scale=0.8]{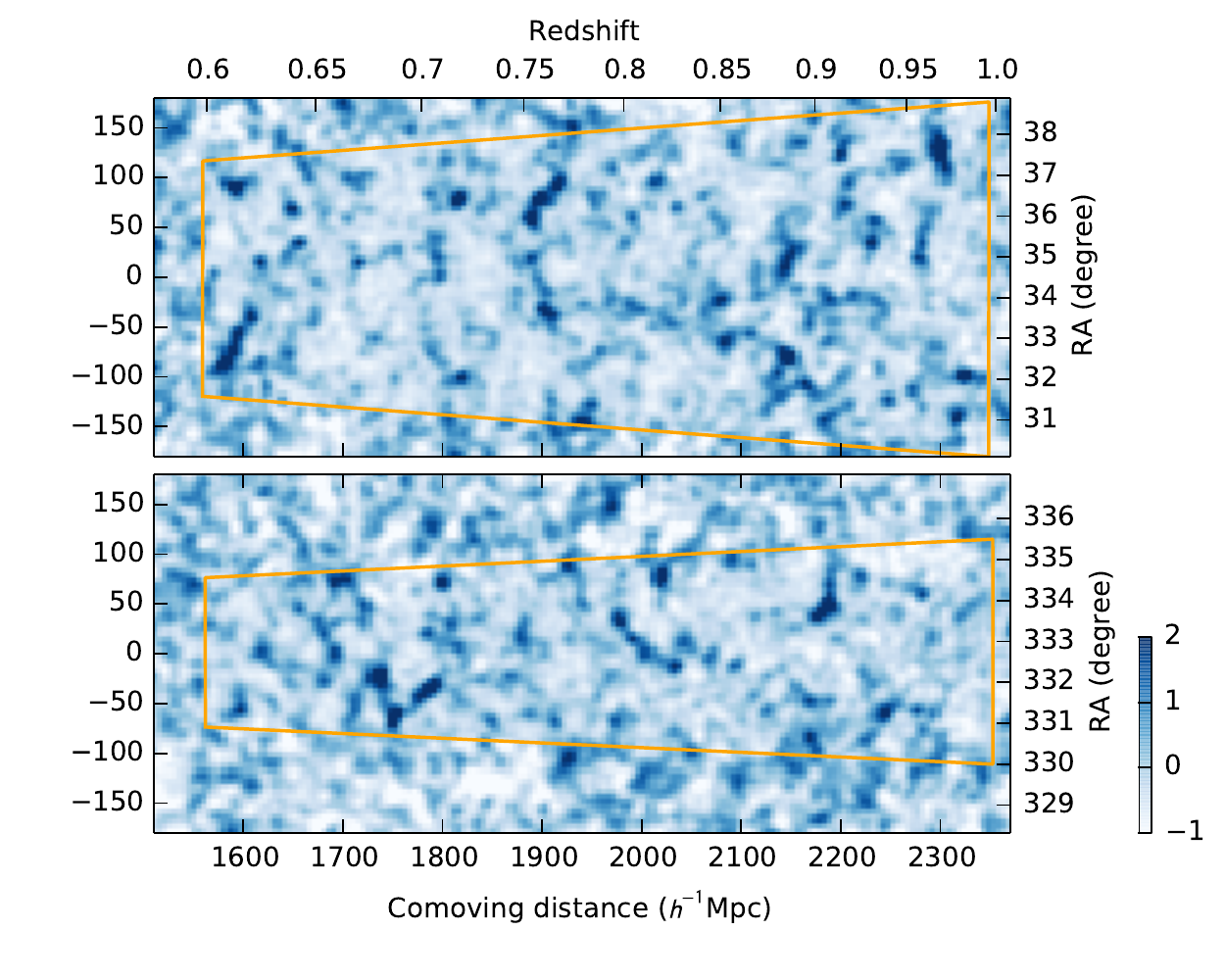}
\caption{VIPERS cone diagrams for the fields  W1 (top) and W4 (bottom).  The left panels show the redshift-space positions of observed galaxies.  The marker colour indicates the blue or red colour class and the marker size scales with B-band luminosity.  The depth of the slice is 10\hmpc.  The orange line traces the field boundaries cut in the redshift direction at $0.6<z<1.0$.  At right we show a slice of the density field taken from one step in the Markov chain.  It represents the anisotropic Wiener reconstruction from the weighted combination of galaxy tracers.  The field is filled with a constrained Gaussian realisation.  The field has been smoothed with a Gaussian kernel with full-width-half-max 10\hmpc.  The colour scale gives the over-density value.
\label{fig:densfield}}
\end{figure*}

\section{Data model}\label{sec:datamodel}
\subsection{Galaxy number counts}
We overlay a three-dimensional cartesian grid on the survey.  The number of galaxies in a given sample observed within a cell indexed by $i$ is
related to an underlying continuous galaxy density field $\delta_{G}$ by
\begin{equation}
N_{i} = \bar{N}  w_{i} (1+\delta_{G,i}) + \epsilon_{i}, \label{eq:countmodel}
\end{equation}
where $w_i$ is the spatial selection function  and $\bar{N}_i$ is the mean density providing the normalisation.  The stochastic nature of galaxy counts is captured by the random variable $\epsilon_i$ and is dominated by Poisson noise except in the highest density peaks \citep{Diporto14}.   The cells are defined in comoving redshift-space coordinates and we adopt a fiducial cosmology to define the relationship between redshift and comoving distance.

The selection function $w$ in Eq. \ref{eq:countmodel} gives the likelihood of observing a galaxy at a given grid point.  It accounts for the angular geometry of the survey,  sampling rate and redshift distribution.  In this 
analysis we separate the angular and line-of-sight components.  As described in Section 5.1, the angular dependence is determined from the survey mask and target sampling rate (TSR) while the redshift distribution is computed assuming the luminosity function and apparent flux limit for the given subsample of galaxies.

The expected number of galaxies in a cell is  given by the product of $\bar{N}$ and the selection function, giving
\begin{equation}
\langle N_{i} \rangle = \bar{N} w_{i} .
\end{equation}

The selection function as described can account for spatial variations but cannot describe sampling dependencies on galaxy type or apparent flux.  For VIPERS data we will up-weight galaxies
based on the inverse spectroscopic success rate (SSR) depending on quadrant and apparent $i$ band magnitude.  These weights are only indirectly correlated with the density field
so they result in an amplification of the shot noise level.  The SSR weight of a galaxy is $w_{SSR}$ and the noise amplification factor is $\alpha=\langle w_{SSR} \rangle$ averaged over all galaxies in the subsample.
Therefore, in Eq. \ref{eq:countmodel} $N_i$ represents the weighted count of galaxies and, consequently, the variance of the stochastic term, $\sigma^2_\epsilon$, is boosted to  $\alpha \sigma^2_\epsilon$.

In this analysis we discretise the galaxy field onto a coarse spatial grid as well as onto a finite grid of Fourier modes.  This process introduces an error in the density field arising from the aliasing of structures:  small-scale structures with spatial frequencies higher than the Nyquist frequency become imprinted on larger scales  \citep{Hockney}.  The effect may be corrected for in the power spectrum by assuming the spectral shape above the Nyquist frequency \citep{Jing05}.  However, to accurately reconstruct the density field without making such assumptions, we may use a mass-assignment scheme or anti-aliasing filter discussed in Appendix~\ref{appendix:aliasing}.  The smoothing effect of mass-assignment schemes introduces a convolution in Eq. \ref{eq:countmodel} which invalidates the simple count model.  An alternative is to use the super-sampling method of  \citet{JascheAliasing} that approximates the ideal anti-aliasing filter and does not damp small-scale power.  Since we desire a compact window in both configuration and Fourier space, we adapt this technique with a soft cut  as described in Appendix~\ref{appendix:aliasing}.  This approach reduces the aliased signal to the level of the triangle-shaped-cell scheme while preserving small-scale power to $k\sim0.7 k_{Nyquist}$.

The convolution introduced by the anti-aliasing filter modifies the noise properties such that the Poisson expectation, $\sigma^2_i=\bar{N}_i w_i$, cannot be assumed.  Instead we  use a re-scaled Poisson variance characterised by the factor $\nu_i\equiv \sigma^2_i/\bar{N}$.  As described in Sec.~\ref{sec:results}, the factor may be estimated in a Monte Carlo fashion given the mask and anti-aliasing filter. 

A cell partially cut by the mask will become strongly coupled with its neighbour through the anti-aliasing filter. In practice we neglect the additional off-diagonal cell-cell contributions in the noise covariance matrix.   However, we found that it is necessary to regularise the noise matrix by increasing the noise level to achieve stable results.  We set a lower limit on the cell variance through a parameter $\nu_{threshold}$ such that the scaled shot noise has a floor set by $\nu_i \leftarrow \max(\nu_{threshold},\nu_i)$.  The procedure is identically applied both to mocks and real data.

\subsection{Galaxy bias}
We assume a constant linear biasing model such that 
\begin{equation}
\delta_{G,i} = b D(z_i)  \delta_i,
\end{equation}
where $z_i$ is the redshift of the cell indexed by $i$.  The bias factor $b$ depends on the luminosity and colour of the galaxy subsample.  We give explicitly the  growth of matter fluctuations with time
according to the linear growth factor $D(z)\equiv\D(z)/\D(z_{ref})$ with $\delta(z) = \delta(z_{ref}) D(z)$.  For the VIPERS sample the reference redshift is set to $z_{ref}=0.7$.  Since 
VIPERS covers an extended redshift range this factor brings large-scale density modes to a common epoch.

\subsection{Number density}
The number density of galaxies in a luminosity bin is given by the integral of the luminosity function:
\begin{equation}\label{eq:lf_int}
n(z) = \int_{M_{bright}}^{M_{faint}(z)} n(M,z) dM.
\end{equation}
We parameterise the luminosity function using the Schechter function \citep{Schechter1976} in terms of magnitudes and the parameters $(\phi_{\star} ,M_{\star} ,\alpha)$:
\begin{equation}
n(M) =0.4 \ln{10} \ \  \phi_{\star}  \left( 10^{0.4\left(M_{\star} - M\right) }\right)^{\alpha + 1} \exp\left( 10^{0.4\left(M_{\star} - M\right)} \right).
\end{equation}
The characteristic magnitude evolves as $M_{\star}(z)=M_{\star}(0) + \E z$ with $\E\approx-1$ for red galaxies in VIPERS \citep{Fritz14} confirming the findings of previous
studies at moderate redshift \citep{Ilbert05,Zucca09}.

The number density observed is further reduced by the survey completeness.  
In VIPERS, galaxies are targeted to an apparent magnitude limit of $m_{lim}=22.5$ in the $i_{AB}$ photometric band.  This sets an absolute magnitude limit for a given class of galaxy
\begin{equation}\label{eq:restframe}
M_{limit}(z) = m_{lim} - D_m(z)  - \K(z),
\end{equation}
where $D_m(z)$ is the distance modulus which depends only on the background cosmology and $\K$ is the K-correction term which depends on the particular type of galaxy targeted.  
The absolute magnitudes were computed for each galaxy by fitting spectral energy distribution templates to broadband photometry as described by \citet{Davidzon13} and from the absolute magnitudes we infer the K-corrections.
We parameterise the trend of K-correction with redshift as $\K(z) = \K_0 + \Delta_K(z)$.  $\K_0$ is estimated from the median value of galaxies within a given subsample while  $\Delta_K(z)$ is a polynomial fit with the following coefficients, different for red and blue galaxy types: 
\begin{eqnarray}
\Delta_{K,blue}(z) = 1.784(z-0.7)^2 + 0.440(z-0.7) -0.678\\
 \Delta_{K,red}(z) = 2.144(z-0.7)^2 + 1.745(z-0.7) - 0.720.
\end{eqnarray}

Using this parameterisation the inferred magnitude limits corresponding to 50\% completeness  are indicated in Fig.~\ref{fig:samples} for the blue and red samples by the solid and dashed lines. 
For mock samples, the K-correction term and its evolution are fixed to $\K_{mock}=z-1.3$.

With these ingredients we model the redshift distribution of each galaxy subsample by integrating the luminosity function with Eqs. \ref{eq:restframe} and \ref{eq:lf_int}.  We let the mean density of each subsample free to set the normalisation of the redshift distribution.  We then take the shape given by the Schechter function to interpolate the luminosity function across the bin.  The parameters $M_{\star}$ and $\alpha$ in each colour and redshift bin are fixed to the values measured in VIPERS \citep{Fritz14}.  Since the precise luminosity evolution is not known, the evolution term, $\E$, is allowed to vary as a function of colour and redshift.  This gives a characteristic magnitude $M_{\star}(z) = M_{\star}(z_{ref}) + \E (z-z_{ref})$ where $z_{ref}$ is taken to be the midpoint of the redshift bin.  Changing $\E$ modifies the shape of the redshift distribution.

\subsection{Power spectrum}
The matter power spectrum in real space $P(k) = \langle |\delta_k|^2 \rangle$ is assumed to be isotropic.  Seen in redshift-space, it is distorted
along the line-of-sight direction \citep{Hamilton98}.  We model the signal on the cartesian Fourier grid as
\begin{equation}\label{eq:pk}
S(k,\mu;\beta,\sigma_v,\sigma_{obs}) =  A \frac{ \left( 1+\beta \mu^2 \right) ^2}{1+k_{los}^2 \sigma_v^2}  e^{{-\frac{k_{los}^2\sigma_{obs}^2}{2}}}  \B^2(k_x,k_y,k_z) P(k),
\end{equation}
where $\mu \equiv k_{los}/k$ and $k = \sqrt{k_x^2+k_y^2+k_z^2}$.  The line-of-sight direction is aligned with the grid such that $k_{los}=k_z$ taking the plane-parallel approximation.  

 The coherent motions of galaxies on large scales are described by the \citet{kaiser87}  factor  with $\beta = f/b_g$ where the growth rate in \LCDM\  is $f(z) = d\log D/d\log a $.
On small scales, velocities randomise and may be modelled by an exponential pairwise velocity dispersion giving a Lorentzian profile in Fourier space which we refer to as the dispersion model  \citep{Ballinger96}.
The velocity dispersion term, $\sigma_v$ in Eq.~\ref{eq:pk}, has units \hmpc. The conversion to velocity units is $H(z)/(1+z)/\sqrt{2}\approx 60.0$\ \mpch\ km s$^{-1}$ which over the redshift range of interest is nearly constant.
We add a Gaussian term along the line of sight to characterise redshift measurement errors where $\sigma_{obs}= \sigma_{cz} /H(z)$ and $\sigma_{cz}$ is the redshift error.  For VIPERS the estimated redshift error is $\sigma_{cz}=141(1+z) $km/s \citep{Guzzo14} and $\sigma_{obs}=1.67$\hmpc\ and is nearly constant over the redshift range 0.6-1.0.

The factor $\B(k_x,k_y,k_z)$ accounts for the cell window function arising from the anti-aliasing filter and is given by Eq. \ref{eq:window}.  In this analysis the absolute amplitude of the power spectrum is not constrained.  So we set the amplitude $A$ in Eq.~\ref{eq:pk} to fix $\sigma_8=0.8$, the variance computed on a scale of $R=8$ \hmpc~integrated to the Nyquist frequency.

We ignore geometric distortions arising from the choice of the fiducial cosmology \citep{Alcock79}.  The resulting bias is not significant when compared with the statistical uncertainties of the VIPERS redshift-space clustering measurements \citep{Delatorre13}.  However, when carrying out a model test, we may rescale the density field and two point statistics to transform from the fiducial to the test cosmology as carried out for the VIPERS power spectrum analysis by Rota et al (in preparation), but this is not done here.

\begin{table}
\begin{tabular}{l|c|l}
\hline
 Parameter             &  Symbol    & Dimension \\
\hline
\multirow{2}{*}{Overdensity field }  & \multirow{2}{*}{$\delta$ }    & $2\times72\times16\times172$ \\
                                                       &                     &(5\hmpc~cubic cells) \\
\hline                                                       
Power spectrum    & $P$            & 109\\
Distortion factor     & $\beta$      & 1 \\
Velocity dispersion & $\sigma_v$& 1 \\
Galaxy bias             & $b$            & 37 (19 blue, 18 red) \\
Mean number density & $\bar{N}$   & 37 \\
Luminosity evolution & $\E$  & 8 \\
\hline
\end{tabular}\caption{Accounting of the free parameters in the data model. \label{tab:params}}
\end{table}

\section{Gibbs sampler}\label{sec:method}
We present a brief overview of the Gibbs sampler.  Since our implementation differs from that of \citet{Jasche13} we provide a detailed description in Appendix~\ref{appendix:method}.  The full parameter set introduced in the previous section is summarised in Table \ref{tab:params}.  We use the Gibbs sampling method to  sample from the joint posterior of the parameter set.  This is performed by iteratively drawing samples from  each conditional probability distribution in the following steps (where \according indicates that a sample is drawn from the given distribution):
 \begin{enumerate}
\item  Generate $\delta^{s+1}$ \according $p(\delta | \bar{N}^s, b^s, P^s,\beta^s,\sigma_v^s,N)$
\item Generate $P^{s+1}$ \according $p(P | \delta^{s+1}, \bar{N}^s, b^s,\beta^s,\sigma_v^s,N)$
\item Normalise power spectrum $P^{s+1}$.
\item Generate $\beta^{s+1}, \sigma_v^{s+1}$ \according $p(\beta,\sigma_v | P^{s+1}, \delta^{s+1},\bar{N}^s,b^s,N)$
\item Generate $\bar{N}^{s+1}$ \according  $p(\bar{N} | b^{s+1},P^{s+1}, \beta^{s+1},\sigma_v^{s+1},\delta^{s+1},N)$
\item Generate $\E^{s+1}$ \according  $p(\E | \bar{N}^s,b^{s+1},P^{s+1}, \beta^{s+1},\sigma_v^{s+1},\delta^{s+1},N)$
\item Generate $b^{s+1}$ \according  $p(b | P^{s+1}, \beta^{s+1},\sigma_v^{s+1}, \delta^{s+1},\bar{N}^{s+1},N)$
\end{enumerate}
These steps are repeated forming a Markov chain and after an initial burn-in period we can expect that the samples are representative of the joint posterior distribution. 

In the first step, we sample from the conditional probability distribution for the density field in a two-stage procedure.  First, the Wiener filter is used to compute the maximum a-priori 
field $\delta_{WF}$ \citep{Kitaura10-LN}.  The Wiener filter solution is a smoothed field that gives an underestimate of the true power.  To generate a realisation of the density field a random component that is uncorrelated with the observations $\delta_{random}$ is added \citep{Jewell04}.   The final field is thus the sum $\delta = \delta_{WF} + \delta_{random}$.  

After constructing a realisation of the density field, the second step is to sample the power spectrum.  We do put a Gaussian prior on the first bin at $k<0.01$\mpch setting the mean and variance to the Fiducial value and sample variance expectation.  This aids the stability of the chain.  A uniform prior is used for the bins at $k>0.01$\mpch.  We use two approaches to sampling the power spectrum detailed in Appendix~\ref{appendix:signal}.  First, we draw samples from the inverse-gamma distribution, see e.g. \citet{Jasche10-SDSSBayes}.  However this produces very small steps in the low signal-to-noise regime and so can be inefficient at small scales.  Therefore, on alternative steps we carry out a Metropolis-Hastings routine to draw samples of the power spectrum according to the likelihood, Eq. \ref{eq:likelihood}.  We find consistent sampling of the power spectrum using the two methods. 

Since we cannot constrain the absolute normalisation of the power spectrum, we normalise to the desired value of $\sigma_8$.  We next draw the redshift-space distortion parameters $\beta,\sigma_v$ which are independent of the power spectrum amplitude.

Next, we sample from the mean density conditional probability distribution for each galaxy sample which includes the evolution factor $\E$.  Here we use a Poisson distribution, as described in  Appendix~\ref{appendix:meandensity}.

Finally we sample from the bias conditional probability distribution for each galaxy sample.  This distribution is  Gaussian for the bias parameter (see Appendix~\ref{appendix:bias}).  In this method, the bias is computed on the redshift-space grid, which in our case has a resolution of 5\hmpc.  For physical interpretation it is interesting to estimate the bias averaged on larger scales.  So, in estimating the bias we first down-grade the grid resolution by a factor of two, such that the bias is averaged over a scale of 10\hmpc.  We impose a uniform prior for the bias values of $0.5<b<4$.

\begin{figure*}
\includegraphics[height=4in]{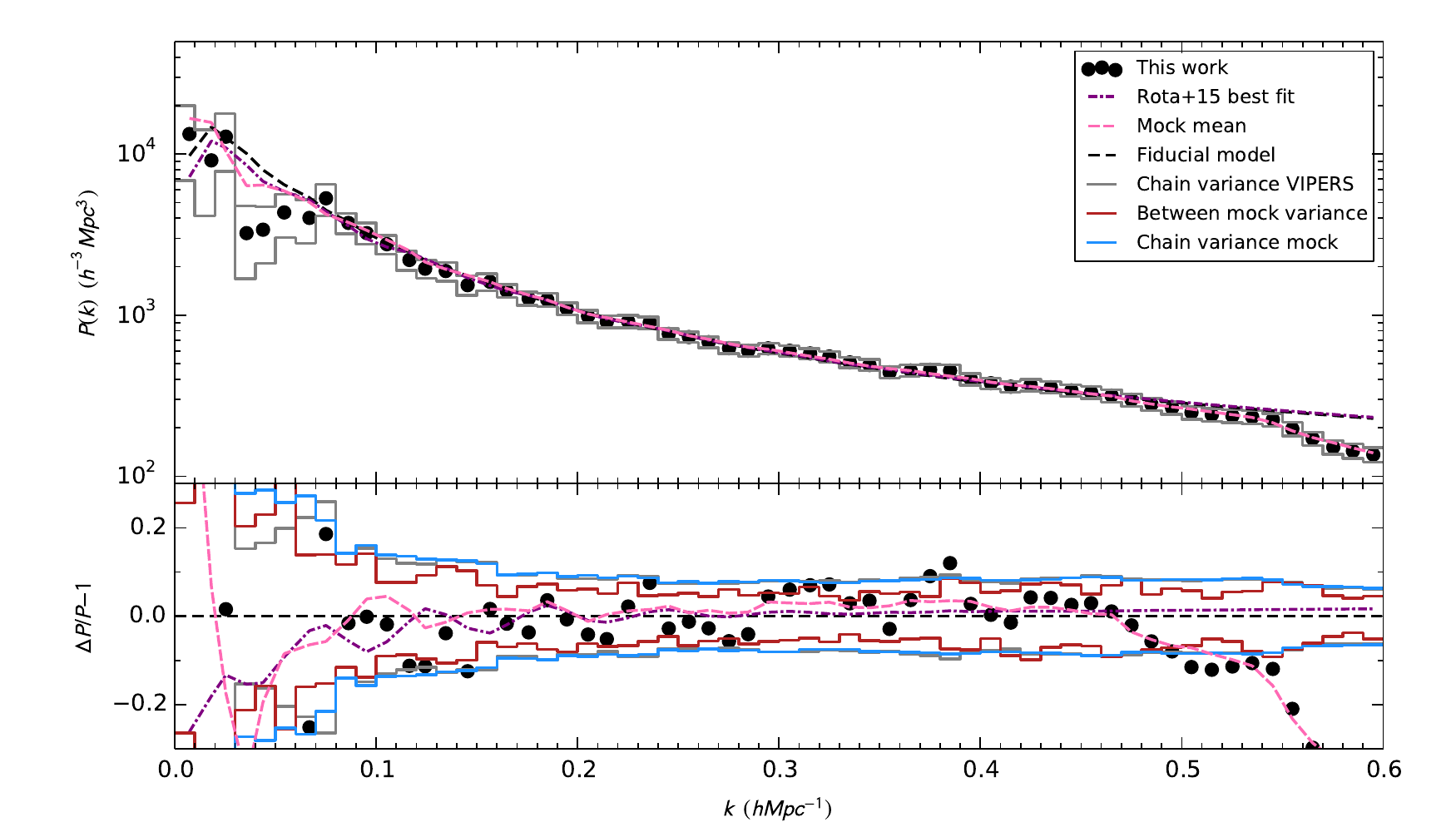}
\caption{Constraints on the real-space power spectrum.  The lower panel shows the relative difference with the fiducial model.   The black dots give our estimates of the binned real-space power spectrum taken from the median of the markov chain.  Overplotted is the fiducial model adopted in this study (black dashed curve).   We find agreement with the best-fit model using VIPERS data by Rota et al (in preparation) (purple dot-dashed curve).   The pink dashed curve is the mean of the power spectrum estimates taken from the 27 mock catalogues.  We present three error estimates: the internal chain variance determined from VIPERS data (grey steps), the chain variance determined from mock catalogues (blue steps) and the variance of the individual estimates from the 27 mock catalogues (red steps). The error corridors show 70\% confidence intervals.   \label{fig:viperspk}}
\end{figure*}

\begin{figure}
\includegraphics[height=3.6in]{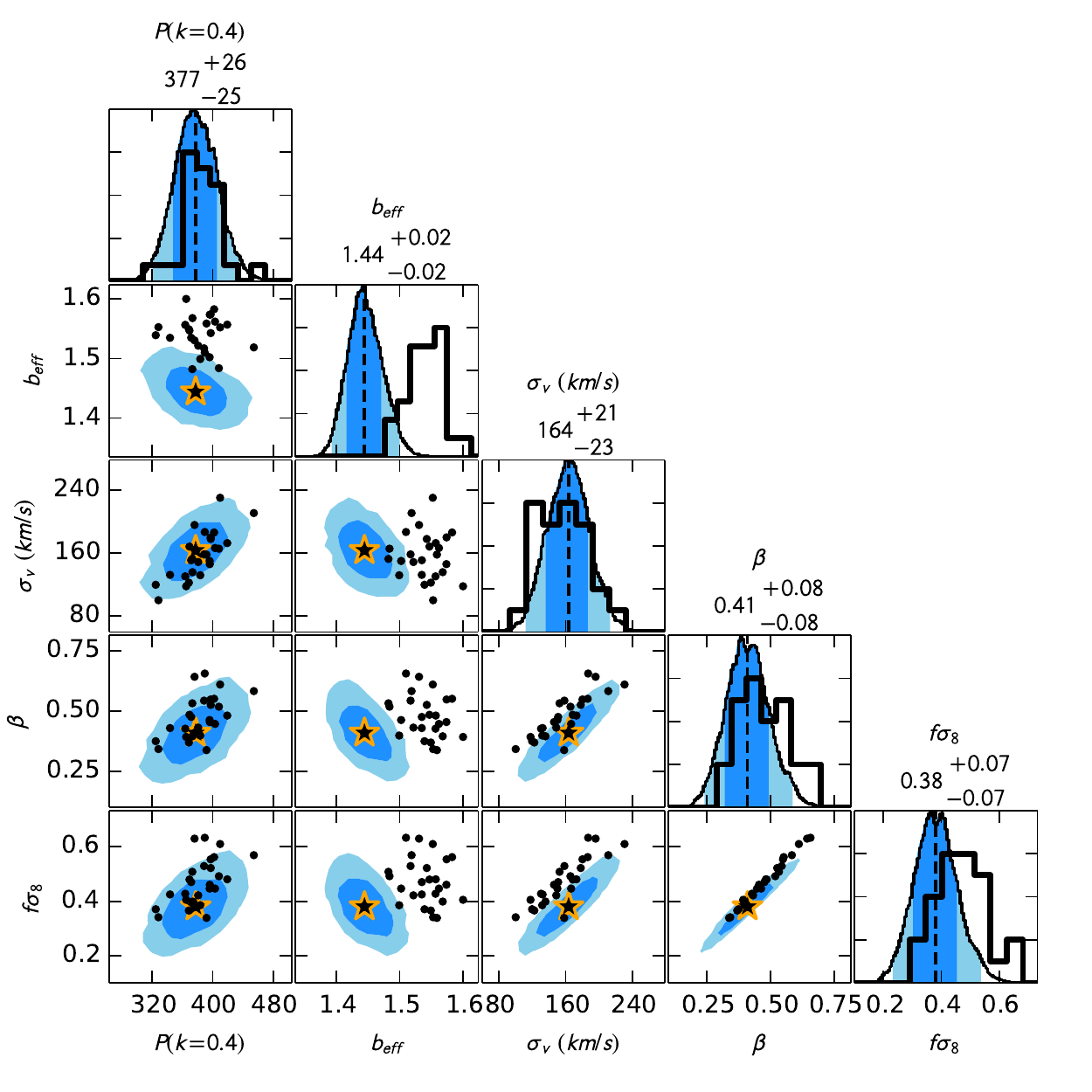}
\caption{ We show the degeneracies between RSD parameters $\beta$,$\sigma_v$, effective bias and the power spectrum at $k=0.4$\mpch.  The shaded regions mark the 68\% and 95\% confidence intervals from VIPERS chain and the star symbols mark the mean value.  The points give the distribution of mean values derived from mock catalogues.  The histograms along the diagonal give the marginalised distributions of each parameter chain.  The filled histogram gives the distribution from the VIPERS chain, while the solid line is the distribution of mean values derived from the mock catalogues. \label{fig:viperspkchain}}
\end{figure}

\section{Application to VIPERS}\label{sec:results}
\subsection{Set-up}
The data and mock catalogues are processed similarly, although the construction of galaxy subsamples differs.  The mock catalogues do not include the inhomogeneous incompleteness corrected for in the data by the SSR  and  TSR factors.  The uncertainties introduced by these corrections are negligible compared with statistical uncertainties in VIPERS. 
\begin{enumerate}
\item The two survey fields, W1 and W4 are separately embedded into rectangular boxes.   The  grids have  dimensions $72\times16\times172$ cubic cells with comoving size 5\hmpc.  We align the grid such that at the field centre, the three axes correspond to the right ascension, declination and line-of-sight directions.  The co-moving coordinates are computed using the fiducial cosmological model.  In the real catalogue only, galaxies are up-weighted by the inverse spectroscopic success rate (SSR) depending on quadrant and apparent $i$-band magnitude.  
\item We compute the density on the grid using  the anti-aliasing filter based on the super-sampling method proposed by \citet{JascheAliasing} with a soft $k$-space cutt-off as described in Appendix~\ref{appendix:aliasing}.   
\item The angular ($\alpha, \delta$) and radial ($z$) components of the selection function are computed separately on the grid: $w(\alpha_i,\delta_i,z_i)=w(\alpha_i,\delta_i)w(z_i)$.  
\item For the angular component, we generate a uniform grid of test points that over-sample the grid by a factor of 8 and reject points outside the survey angular mask.  For VIPERS data the remaining are down-sampled by the TSR.  The points  are then assigned to the grid points using the anti-aliasing mass-assignment scheme.  The selection function $w_i$ is then given by the normalised density of test points on the grid.
\item The radial component of the selection function is estimated in bins of redshift, luminosity and colour.  We estimate the median $K$-correction term for galaxies within each bin and use Eq. \ref{eq:lf_int} to compute the unnormalised $N(z)$.
\item We  estimate the generalised shot noise variance  $\nu_i \equiv \sigma^2_i/\bar{N}_i$ which depends on the mask through the  anti-aliasing filter in Monte Carlo fashion.  We generate a set of 1000 shot noise maps by distributing random points over the survey volume.  For VIPERS data the points are down-sampled by TSR.  We then compute the variance for each cell of the map over the 1000 realisations.  To regularise the noise covariance matrix, a threshold is set $\nu_i = \max(\nu_{thresh},\nu_i)$ where $\nu_{thresh}=0.3$ for mocks and $0.15$ for data to account for TSR.
\end{enumerate}

With the galaxy number density map, selection function and noise map, we have all the components of the data model required to estimate the posterior probability distribution in Eq. \ref{eq:bayes}.  To sample this probability distribution we run the Gibbs sampler Markov chain for 2000 steps and, allowing for a burn-in period, begin the analysis from step 1000.  The convergence properties and justification for the burn-in period are shown in Appendix \ref{appendix:conv}.  For the VIPERS data we ran 7 independent chains for 2000 step each providing 7000 post-burn-in samples for analysis.

Taking the variance of the Markov chain gives us an internal error estimate on the parameters.  The runs on mock catalogues show that the chain variance corresponds to the expected sample variance for the power spectrum.  However, this is not necessarily true for the other statistics.  For instance the luminosity function quantifies the distribution of observed galaxies which remains fixed in the chain.  It is only indirectly dependent on the underlying density.

\subsection{Density field}
The density field taken from a single step (1500) in the Markov chain  is shown in Fig.~\ref{fig:densfield}.  It represents the application of the Wiener filter on a bias-weighted combination of the galaxy subsamples.  The reconstruction is based on the redshift-space distortion model and so the resulting field is anisotropic and characterised by effective redshift-space distortion parameters averaged over the galaxy samples.

The result of the Wiener filter is an adaptively smoothed field that extrapolates structures over the correlation length of a few megaparsecs.  To build a full realisation of the structures we add a Gaussian constrained realisation that fills in the gaps and gives the full variance.  The structures outside the survey boundary are generated from a random Gaussian realisation although the phases are properly aligned at the boundary.  The true galaxy density field on scales of 5\mpch\ is far from Gaussian and the difference  is visible by eye. 

In Fig.~\ref{fig:densfield} we can recognise the cosmic web of structures including knots, filaments and void regions.  The structures are richest where the sampling is highest at lower redshift.  At redshift $z>0.8$ we see fewer coherent structures and  the contribution from the constrained Gaussian realisation is larger.  Each step of the Markov chain gives a  reconstruction of the field with different realisations of noise and large-scale modes.  Once the chain has passed the burn-in period (see Appendix \ref{appendix:conv}), we can consider these realisations to represent Gaussian perturbations around the observed galaxy field.

\subsection{Redshift-space power spectrum}
The galaxy power spectrum in redshift space is parameterised in terms of the real-space matter power spectrum, bias and redshift-space distortion factors (Eq. \ref{eq:pk}).  We bin the power spectrum  linearly with bin size $\Delta k=0.01$ giving 109 bins. The redshift-space distortion parameters are fit to $k<0.4$\mpch.  This limit corresponds to $k\sigma_v\approx1$ where we can expect the dispersion model to break down.

The Markov chain provides joint samples of the parameters.  In Fig.~\ref{fig:viperspk} we show the median over the power spectrum chain (black dots).   The confidence coordior gives the 1-$\sigma$ confidence interval estimated from the chain variance (grey steps).  We find good agreement with the model  computed with CLASS and Halofit  (black dashed) with  $\Omega_m=0.27$ \citep{CLASS,halofit1,halofit2}.  The best-fitting model determined by Rota et al (in preparation) has $\Omega_m=0.272 \pm .03$ (over-plotted with purple dot-dashed curve).

 On small scales $k>0.45$\mpch\ ($0.75\times$ Nyquist frequency) the power drops.  This is due to neglecting correlations between cells that arise due to the anti-aliasing filter.  On large scales, there is a dip in power at $k=0.05$\mpch seen in both mock catalogues and data, although it biases the estimate only at the 1$\sigma$ level.  Scales at $k<0.06$\mpch\ are only measured in the line-of-sight direction with VIPERS so the inability to reconstruct them properly without a prior constraint is not surprising.

The median values for the redshift-space distortion parameters we find are  $\beta_{VIPERS}=0.41$ and $\beta_{mock}=0.47$.  The within-chain variance is $\sigma_{chain}=(-0.12,+0.10)$ (68\% confidence interval), while the scatter of the 26 mocks gives standard deviation $\sigma_{mock}=0.09$.  

We may compute the growth rate  through the relation
\begin{equation}
f\sigma_8(z)=\beta \sigma_{8,galaxy}(z),
\end{equation}
where  $\sigma_{8,galaxy}=b_{eff}\sigma_{8}$.  In this analysis we have fixed the amplitude of the matter power spectrum at redshift $z=0.7$  with $\sigma_8(z=0.7)=0.643$ which corresponds to $\sigma_8=0.8$ at $z=0$ in the fiducial cosmology.  

We compute the effective galaxy bias as the number-weighted average over the galaxy samples,
\begin{equation}
b_{eff}^2 = \frac{\sum_{l,i} \bar{N}_l w_{l,i} b_l^2}{\sum_{l,i} \bar{N}_l w_{l,i}},
\end{equation}
where  the sums are over the galaxy subsamples and selection function grids.

We summarise the constraints on the growth rate in Table \ref{tab:RSD}.  We find $(f\sigma_8)_{VIPERS}=0.38_{ -0.07}^{+0.06}$ and $(f\sigma_8)_{mock}=0.46_{-0.12}^{+0.09}$ at $z=0.7$ where we quote the chain variance.  The scatter between the mocks gives a standard deviation $\sigma_{f,mock}=0.08$.  Thus these constraints on the growth rate  are in agreement with the VIPERS correlation function measurement by \citet{Delatorre13}.  There, the error was 16\% on the growth rate $f\sigma_8=0.48$ at $z=0.8$.  In this work we find an error of 18\%.   We attribute the higher error in this analysis to the fact that we marginalise over the real-space power spectrum, while in the previous analysis it was fixed to a fiducial cosmology. 

The correlations between a subset of the power spectrum and redshift-space distortion parameters are shown in Fig. \ref{fig:viperspkchain}.  The star symbols mark the median values of the parameters estimated from the VIPERS Markov chain while the filled contours give the 70\% and 90\% confidence intervals.  The median value and marginalised 70\% uncertainty on each parameter are labeled.  We find that  $f\sigma_8$ (18\% relative error) is better constrained than $\beta$ (20\% error).  This is due to the anti-correlation between $b_{eff}$ and $\beta$ with correlation coefficient $\rho=-0.58$.  These correlations arise from the specific parameterisation adopted and would be modified under a different data model.

The black dots in Fig. \ref{fig:viperspkchain} represent the median values estimated from individual mock catalogues.   We find that the value of the galaxy bias is different within the mocks ($b_{eff}=1.55$) and VIPERS  ($b_{eff}=1.44$).  The bias of the mock galaxies is determined by the luminosity-dependent HOD prescription \citep{Delatorre13} and so the minor difference from real data is not unexpected.  Accounting for the difference in bias, we find excellent agreement between the distribution of mocks and the parameters estimated from real data.  The similarity of the PDF shapes also gives us confidence in the analysis method and error estimates.  Furthermore, since the mocks do not include many of the selection effects present in the data the agreement suggests that these sources of systematic uncertainties are not influencing our conclusions.

\begin{table*}
\begin{tabular}{c|rrr|rrr|rrr}
\hline
          & $\beta$ & $\sigma_{\beta,chain}$ &$\sigma_{\beta,mock}$  &    $b_{eff}$ & $\sigma_{b,chain}$ &$\sigma_{b,mock}$ &   $f\sigma_8$ & $\sigma_{f,chain}$ & $\sigma_{f,mock}$\\
\hline
Mock & $0.47$ & $-0.12$/$+0.09$ & $0.09$ & $1.54$ & $-0.04$/$+0.04$ & $0.03$ & $0.46$ & $+0.09$/$-0.12$ & $0.08$ \\
\hline
VIPERS & $0.41$ & $-0.08$/$+0.07$ &  & $1.44$ & $-0.03$/$+0.02$ &  & $0.38$ & $-0.07$/$+0.06$ &  \\
\hline
\end{tabular}\caption{The constraints on redshift-space distortion parameters.  We give the 68\% confidence intervals from the chains and the standard deviation amongst the 26 mock catalogues.  The fiducial value is $f\sigma_8(z=0.7)=0.45$. \label{tab:RSD}}
\end{table*}

\subsection{Colour and luminosity dependent galaxy bias}
We compute the galaxy bias from the variance of the galaxy counts on the grid.  However, we first down-sample the grid by a factor of two such that the bias is computed on a scale of 10\hmpc.  

\begin{figure}
\includegraphics[width=3.5in]{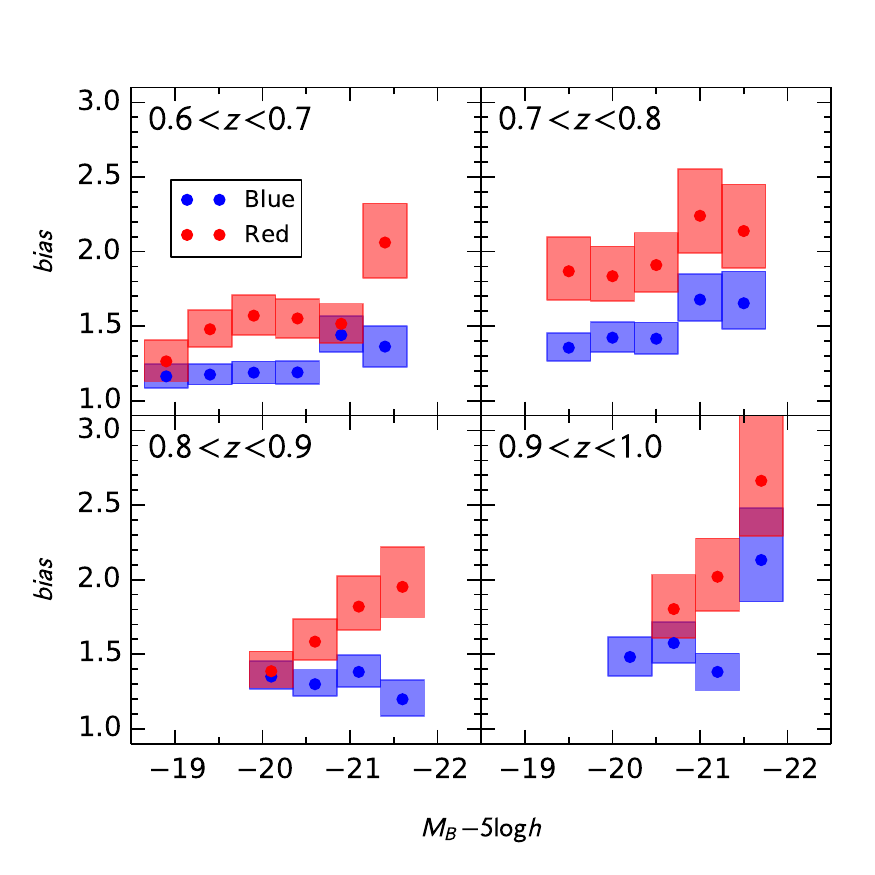}

\caption{The VIPERS galaxy bias parameters in redshift, luminosity and colour bins.  A colour bimodality is seen in each redshift bin.  The trend with luminosity is most striking in the lower redshift bins for both blue and red galaxies. \label{fig:bias}}
\end{figure}

In Fig.~\ref{fig:bias} we show the median bias values of the Markov chain and the confidence intervals are given by the chain variance.  The bias is computed in bins of redshift, luminosity and colour.    
We find a colour bimodality with red galaxies more strongly biased than blue.  This corresponds to the well known galaxy morphology-density relationship that early type galaxies are predominantly found in high density environments \citep[\eg][]{Cucciati06,Dressler80,davis76}.  Similarly, we expect to find that galaxy bias increases with galaxy luminosity since more massive and more luminous galaxies tend to form in more massive dark matter clumps \citep{Coupon12}.  

\begin{figure*}
\center{\includegraphics[width=6in]{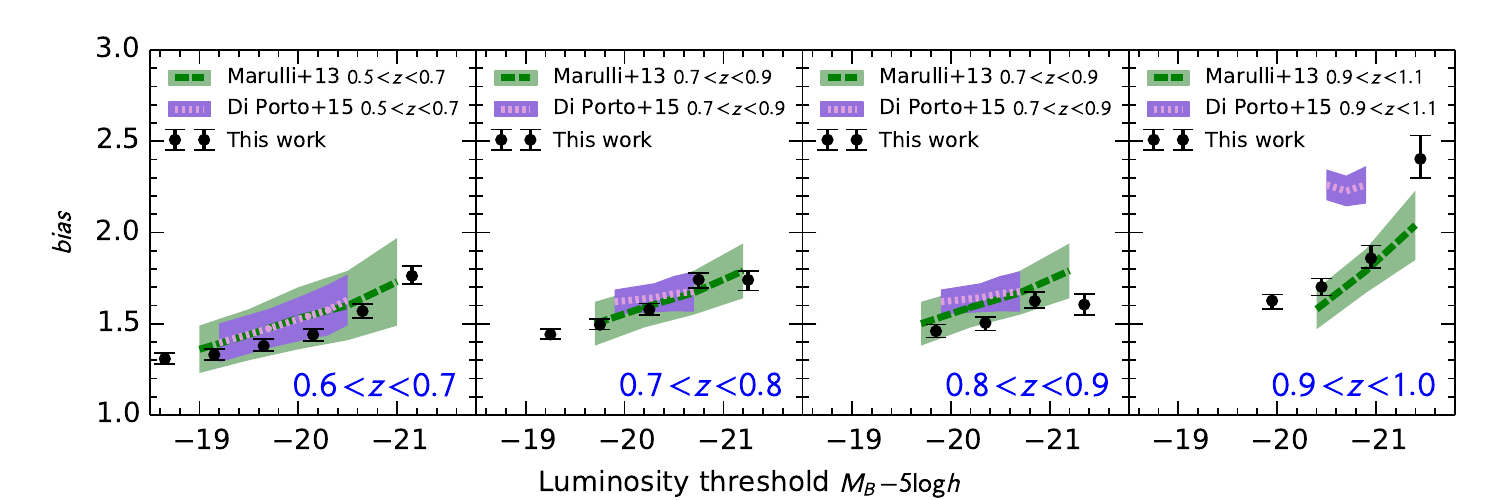}}
\caption{The galaxy bias measured from the full (red and blue combined) VIPERS galaxy sample in luminosity threshold bins.  Reference data are taken from the VIPERS projected correlation function analysis \citep{Marulli13}  and counts-in-cells probability distribution function analysis \citep{Diporto14}.  Note that the redshift ranges differ.
\label{fig:bias_ref}}
\end{figure*}

Previous studies with VIPERS data estimated the galaxy bias of the full galaxy sample as a function of luminosity and redshift.  \citet{Marulli13}  measured the projected galaxy correlation function in luminosity and redshift bins to constrain the mean bias averaged over scales $5-20$\hmpc.  \citet{Diporto14} modeled the counts-in-cells probability distribution function to estimate the linear bias.
 To compare our estimate of the galaxy bias with these previous results we construct luminosity threshold samples counting both red and blue galaxies.  We cross-correlate these number density maps with the Wiener density field from the core analysis and estimate the measurement uncertainty from the chain variance.   The resulting bias values are shown by the markers with error bars in Fig. \ref{fig:bias_ref}.  We find excellent agreement with the previous analyses, although our redshift bins differ.  The bias values from \citet{Diporto14} have been taken on a scale $R = 8$\hmpc~ while those of \citet{Marulli13} are  sensitive to smaller scales. The disagreement at $z>0.9$  may  indicate that the bias of luminous galaxies is scale dependent at high redshifts probed differently by the three studies.

\subsection{Luminosity function}
\begin{figure}
\includegraphics[width=3.5in]{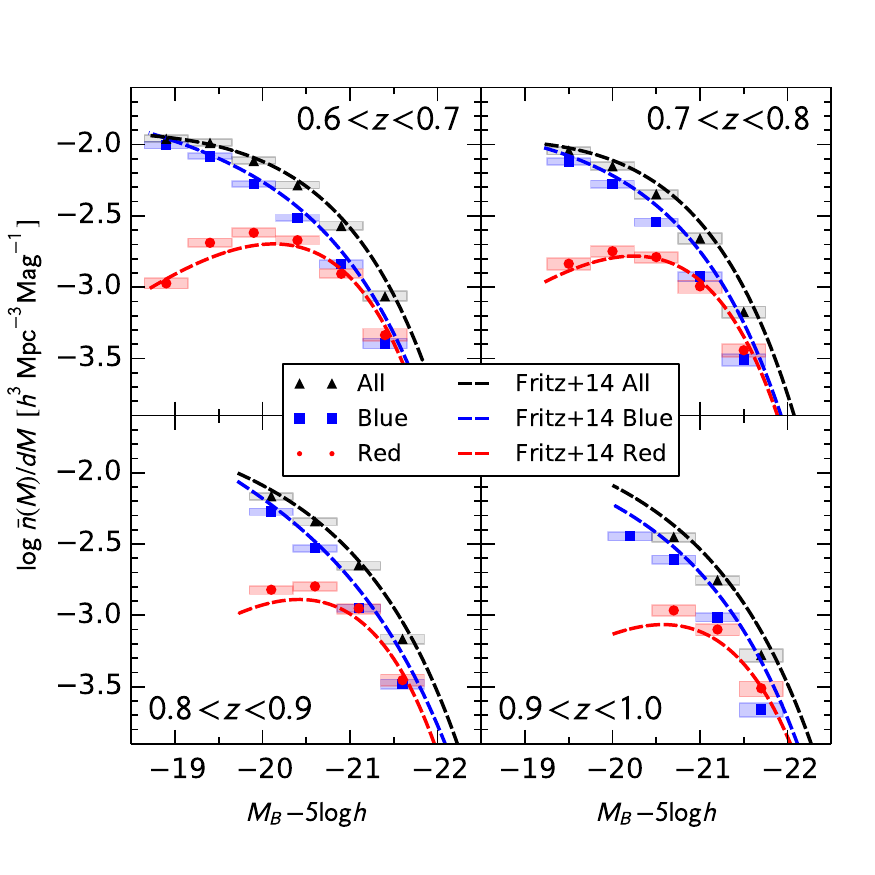}

\caption{The galaxy luminosity function inferred from the mean density Markov chain for red, blue and combined samples in redshift bins.    Markers are plotted at the median value of the chain and the height of the rectangles indicates the 68\% confidence interval. The Schechter function fits from \citet{Fritz14} are overplotted for comparison.   \label{fig:lumfunc}}
\end{figure}

In Fig.~\ref{fig:lumfunc} we show the derived luminosity function based on the mean galaxy number density (dots with error bars) for different galaxy types.  We compare the result to the analysis from \citet{Fritz14} based on the STY \citep{Ilbert05,Sandage79} estimator (dashed curves).  
We can expect to find a difference in the two estimates arising from how the galaxy bias is treated.  The STY estimate is designed to be independent of the underlying density field under the approximation that the galaxy luminosity is uncorrelated with density.   A strong luminosity dependence of the bias can systematically tilt the inferred luminosity function \citep{Smith12,Cole11}.
The agreement between our analysis and the STY measurement indicates that the VIPERS volume is large enough such that sample variance does not significantly alter the amplitude and that any effects due to the luminosity dependence of bias are weak.

\begin{figure}
\includegraphics[scale=1.0]{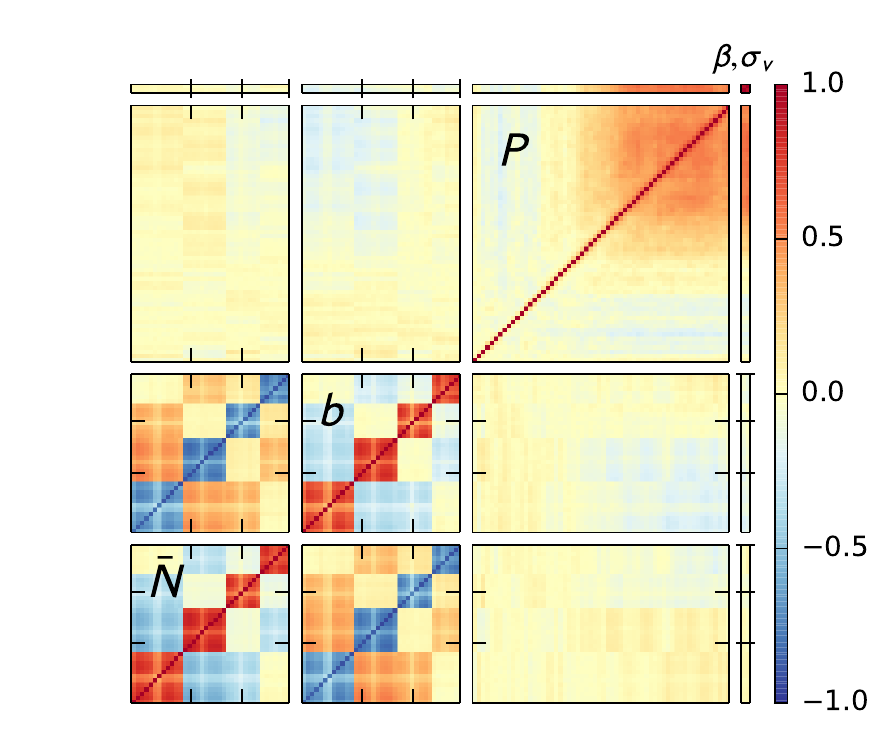}
\caption{The normalised correlation matrix of the parameters computed from the VIPERS Markov chain.  The blocks represent the mean density,  galaxy bias, power spectrum and RSD parameters.   The structure in the covariance arises from the data model parameterisation.  The values of luminosity and colour dependencies of galaxy bias and mean density within a redshift bin are strongly correlated, while they are only weakly correlated across redshift.  On large scales the power spectrum covariance is diagonal, but at $k>0.3$\mpch the bins become correlated due to coupling of the small-scale power with the redshift-space distortion parameters.   \label{fig:cov}}
\end{figure}

\subsection{Parameter covariance}
We estimate the covariance of the statistics with the Markov chain.  Fig. \ref{fig:cov} shows the normalised correlation matrix determined in our analysis,
\begin{equation}
\rho^2_{ij} = \frac{C_{ij}}{\sigma_i \sigma_j}.
\end{equation}
The bias and mean number density parameters are ordered first by luminosity and colour and then by redshift bin.  The appearance 
of blocks in the matrix indicates that within redshift bins the statistics are strongly correlated.  We also find that bias and mean density are anti-correlated, that is, increasing bias necessitates decreasing mean density to preserve the same fluctuation.  The bias and mean density parameters are weakly correlated with the power spectrum measurement.  

The bins of the power spectrum (spacing $\Delta k=0.01$\mpch) show independence on large scales, as is expected for the Gaussian data model, but they become correlated at $k>0.3$\mpch.  The correlations arise from the redshift-space parametrisation that couples the amplitude of the power spectrum to $\beta$ and $\sigma_v$.  The upper right square in the figure represents the nearly 100\% correlation between these two parameters.

\section{Conclusions}\label{sec:conclusions}
Using VIPERS we have demonstrated a method to reconstruct the galaxy density field jointly with the redshift-space power spectrum, galaxy biasing function and galaxy luminosity function with minimal priors on these parameters.  The Bayesian framework naturally accounts for the correlations between these observables.  We adopt a likelihood function for the galaxy number counts that is given by a multivariate Gaussian and set a Gaussian prior on the density field.  The solution that maximises the posterior distribution is given by the classical Wiener filter.  To sample from the posterior distribution we add a Gaussian constrained realisation.  Incorporating this density field estimator within a Gibbs sampler, we jointly sample the full posterior distribution including the power spectrum, bias and luminosity function parameters.
We find encouraging results using a multivariate Gaussian model for the likelihood and prior distributions, although more theoretically motivated descriptions may be used \citep{Kitaura12-2MASS,Jasche13-IC}.

There are clear gains when jointly estimating correlated parameters.  For instance the galaxy colour-density relation can be used to improve estimates of the density field.  Furthermore it is well known that bias weighting galaxies when estimating the power spectrum leads to improved accuracy \citep{Percival04} and greater statistical power \citep{Cai11}.

Moreover, the Bayesian framework provides a recipe for propagating uncertainties in the measurement, incorporating prior knowledge and constraints from the data and guarantees reliable error estimates.
 In VIPERS we account for inhomogeneous sampling and detailed angular masks.  We correct for the selection function of VIPERS by up weighting galaxies according to the magnitude-dependent spectroscopic sampling rate (SSR), while including the target sampling rate (TSR) in the angular dependence of the survey selection function.  These corrections are fixed in our analysis, although for upcoming surveys it will be important to propagate the uncertainties in the selection function to the data products.  

Investigating the covariances between parameters, we find strong correlations between galaxy bias and number density parameters within a given redshift bin.  This is not unexpected since both these parameters depend on the one-point probability distribution function of the density field.   On the other hand the correlation with the power spectrum is weak.  

Our estimate of the power spectrum is effectively deconvolved from the survey window function (see Rota et al. {\it in prep.}) and we find that the covariance of the power spectrum bins is diagonal on large scales as expected from an unmasked Gaussian random field.
On small scales, $k>0.3$\mpch we find significant correlations between power spectrum bins.  On these scales correlations are expected due to the physical processes of structure formation; however, in this case the correlations arise from the parameterisation of the data model.  There is a degeneracy between the redshift-space distortion factors $\beta$ and $\sigma_v$ and the amplitude of the power spectrum on small scales. 
Nevertheless, the error estimate given by the Gibbs sampler matches well the expectation of cosmic variance estimated from mock catalogues.

Our results are in good agreement with previous VIPERS measurements.  We find values of the redshift-space distortion factor $\beta$ that are consistent with the correlation function analysis \citep{Delatorre13}.    The values of luminosity dependent bias we find follow the trends expected from \citet{Marulli13} and \citet{Diporto14} at $z<0.9$.  We further estimate the galaxy bias for colour samples finding a more pronounced dependence on luminosity for red galaxies than blue.  The luminosity function we infer from the mean number density matches well with those found by \citet{Fritz14} using the STY estimator. 

In our analysis we have left the power spectrum, galaxy bias and number density without parameterisation.  Despite this freedom, we find that the resulting error in the key quantities such as the distortion parameter are only marginally larger than that given by traditional methods which can be strongly dependent on parameterisation.   

Our methodology may be extended to jointly analyse multiple datasets in a self-consistent manner.  A particular challenge when considering multiple surveys is dealing with the differences in angular coverage,  sampling rates and galaxy types.   The Bayesian approach provides a means to homogenise datasets allowing for consistent measurements.  

For future studies with VIPERS we can consider the joint analysis with the VVDS-Wide spectroscopic survey \citep{VVDSwide}.  Although the two surveys partially overlap, the selection function and sampling rates differ prohibiting their direct combination.  However, through the Bayesian framework, the joint analysis becomes natural.  We may further add constraints given by the density reconstructions in the gaps by the ZADE algorithm \citep{Cucciati14} or photometric redshift samples from the full CFHTLS Wide fields \citep{Granett12,Coupon12}.  Sheer measurements in these fields can provide additional constraints on the underlying matter density providing a powerful probe in combination \citep{Coupon15}.  
For upcoming surveys, this strategy will guarantee a complete and self-consistent picture of the Universe.

\begin{acknowledgements}
This work is based on observations collected at the European Southern
Observatory, Cerro Paranal, Chile, using the Very Large Telescope
under programs 182.A-0886 and partly 070.A-9007.  Also based on
observations obtained with MegaPrime/MegaCam, a joint project of CFHT
and CEA/DAPNIA, at the Canada-France-Hawaii Telescope (CFHT), which is
operated by the National Research Council (NRC) of Canada, the
Institut National des Sciences de l'Univers of the Centre National de
la Recherche Scientifique (CNRS) of France, and the University of
Hawaii. This work is based in part on data products produced at
TERAPIX and the Canadian Astronomy Data Centre as part of the
Canada-France-Hawaii Telescope Legacy Survey, a collaborative project
of NRC and CNRS. The VIPERS web site is http://www.vipers.inaf.it/.

We acknowledge the crucial contribution of the ESO staff for the
management of service observations. In particular, we are deeply
grateful to M. Hilker for his constant help and support of this
program. Italian participation to VIPERS has been funded by INAF
through PRIN 2008 and 2010 programs.  LG, and BRG acknowledge
support of the European Research Council through the Darklight ERC
Advanced Research Grant (\# 291521).  AP, KM, and JK have been
supported by the National Science Centre (grants
UMO-2012/07/B/ST9/04425 and UMO-2013/09/D/ST9/04030), the Polish-Swiss
Astro Project (co-financed by a grant from Switzerland, through the
Swiss Contribution to the enlarged European Union), and the European
Associated Laboratory Astrophysics Poland-France HECOLS.  KM was
supported by the Strategic Young Researcher Overseas Visits Program
for Accelerating Brain Circulation (\# R2405).  OLF acknowledges
support of the European Research Council through the EARLY ERC
Advanced Research Grant (\# 268107).  GDL acknowledges financial
support from the European Research Council under the European
Community's Seventh Framework Programme (FP7/2007-2013)/ERC grant
agreement \# 202781. WJP and RT acknowledge financial support from the
European Research Council under the European Community's Seventh
Framework Programme (FP7/2007-2013)/ERC grant agreement \# 202686. WJP
is also grateful for support from the UK Science and Technology
Facilities Council through the grant ST/I001204/1. EB, FM and LM
acknowledge the support from grants ASI-INAF I/023/12/0 and PRIN MIUR
2010-2011. LM also acknowledges financial support from PRIN INAF
2012. YM acknowledges support from CNRS/INSU (Institut National des
Sciences de l'Univers) and the Programme National Galaxies et
Cosmologie (PNCG). CM is grateful for support from specific project
funding of the {\it Institut Universitaire de France} and the LABEX
OCEVU.
\end{acknowledgements}

\bibliographystyle{aa}
\bibliography{vipers,surveys,lumfunc,powerspec,mcmc,wiener,voids,galev,theory}

\appendix
\section{Anti-aliasing filter}\label{appendix:aliasing}
\begin{figure}
\includegraphics[scale=0.9]{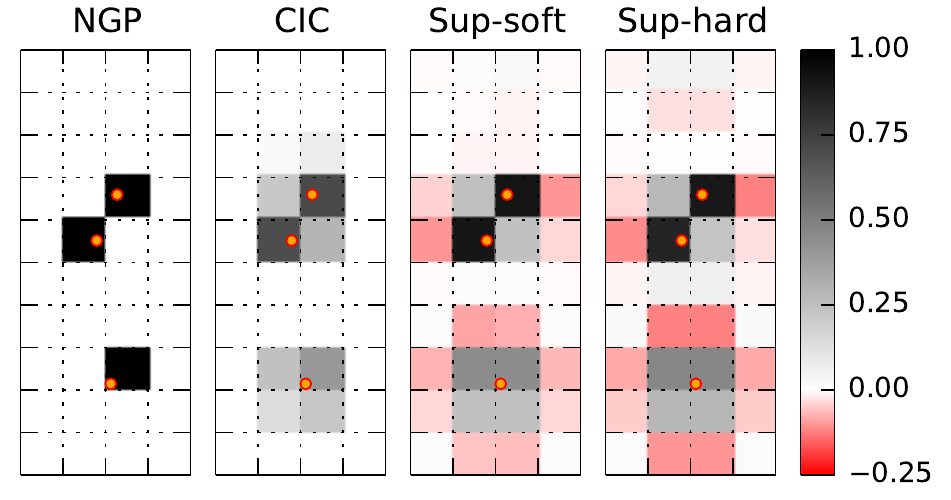}
\includegraphics[scale=0.9]{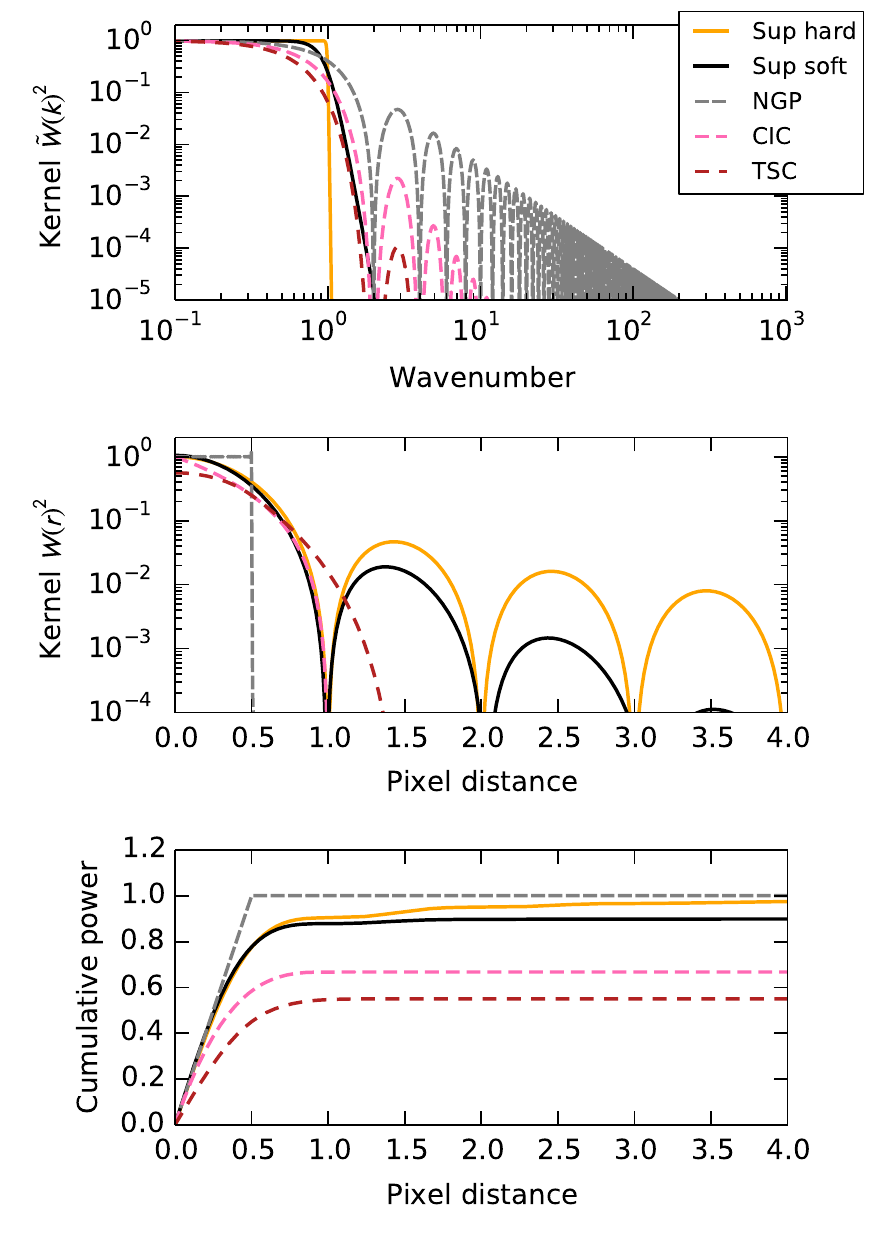}
\caption{Top panel: an illustration of the mass-assignment kernels in two-dimensions: a hard k-space cut ({\it Sup hard}), a soft k-space cut ({\it Sup soft}), nearest-grid-point ({\it NGP}) and cloud-in-cell ({\it CIC}).  The lower three panels give a comparison of the kernel functions (in one dimension) for various mass assignment schemes.  Top: the kernels in Fourier space.  The Nyquist frequency is 1 pixel$^{-1}$ on the scale.  Middle: the kernels in position space.  Bottom: the cumulative power of each kernel in position space.    \label{fig:kern}}
\end{figure}

\begin{figure}
\includegraphics[scale=0.9]{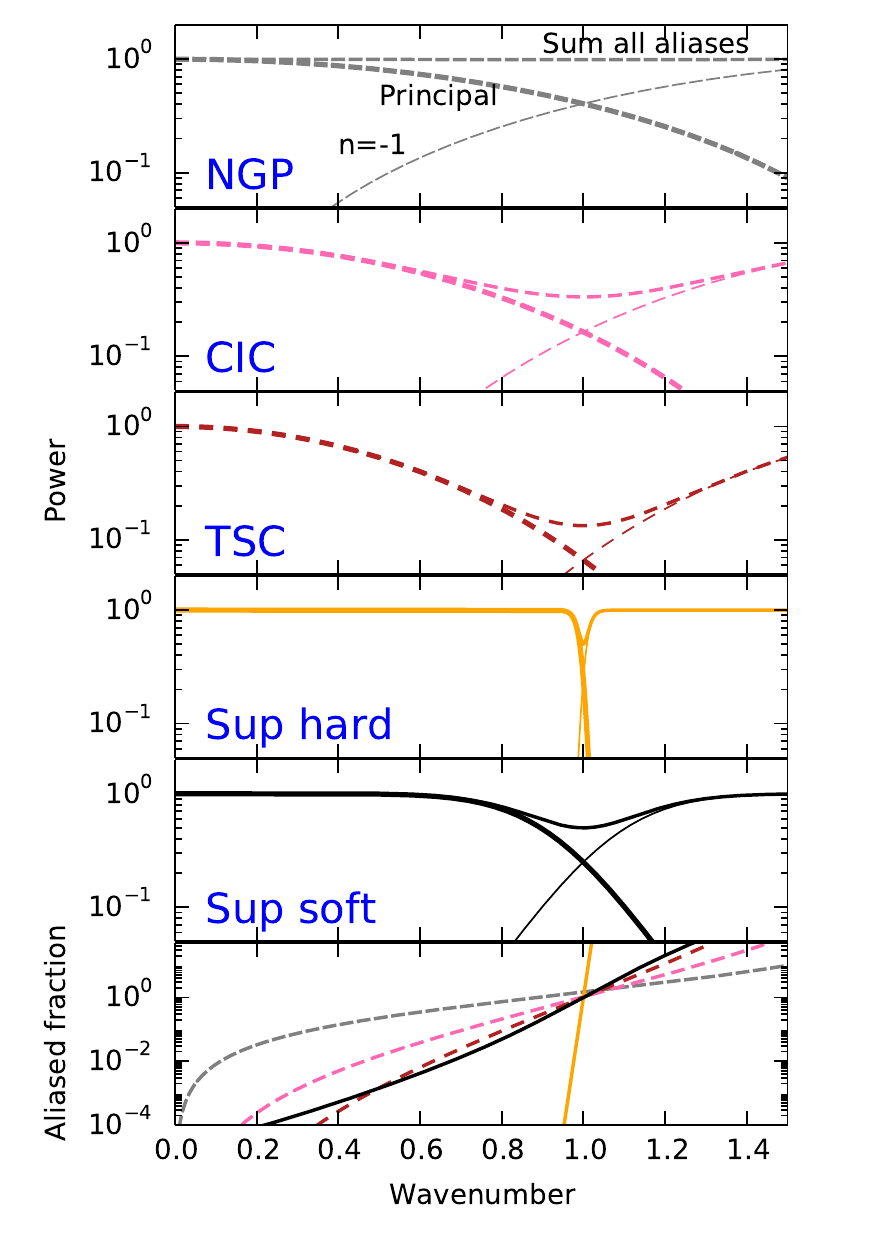}
\caption{A comparison of the shot noise power as a function of wavenumber under various mass assignment schemes.  The Nyquist frequency is 1 pixel$^{-1}$ on the scale.  
The top four plots show the principal component (thick line), the $n=-1$  harmonic and the sum of all harmonics  (thin line).  The bottom frame shows the
fraction of aliased power for each assignment scheme.\label{fig:alias}}
\end{figure}
Binning the continuous galaxy density field onto a grid may be characterised as a smoothing operation followed by discrete sampling \citep{Hockney}.  In the simplest approach, if particles are assigned to the nearest grid point (NGP) the smoothing kernel $W(x)$ has a top-hat shape with the dimension of the grid cell.   More extended kernels can be chosen that distribute the weight of the particle over multiple cells.  Common assignment schemes include cloud-in-cell (CIC) and triangle-shaped-cell (TSC) which correspond to iterative smoothing operations with the same top-hat filter.  Smoothing the field damps small scale power, but better localises the signal in k-space which is beneficial in Fourier analyses.

A consequence of using the fast Fourier transform (FFT) algorithm is that the signal must be discretised onto a finite number of wave modes or equivalently that periodic boundary conditions are imposed.
The observed power is thus a sum of all harmonics of the fundamental wavelength which are known as aliases \citep{Jing05,Hockney}:
\begin{equation}
P'(k) = \sum_n |W(k+2k_N n)|^2 P(k+2k_N n)
\end{equation}
As long as the signal is sufficiently well sampled and band-limited, meaning that there is no power above the Nyquist frequency, the signal may be recovered without error.  However, most signals of interest are not band limited, and so after sampling onto the grid, the harmonics overlap and the signal cannot be exactly recovered.   

The ideal anti-aliasing filter in k-space  is the  top-hat that cuts all power above the Nyquist frequency.  However, the signal cannot be localised both in k-space and in position-space and a sharp cut in k-space will distribute the mass of a particle over the entire grid.  

A practical implementation of the ideal filter was developed by \citet{JascheAliasing}.  The method involves first super-sampling the field by assigning particles to a grid with resolution higher than the target grid.  Transforming to Fourier space via FFT, the high resolution grid is filtered setting to 0 all modes above the target Nyquist frequency.  Finally the grid is transformed back to position-space and down sampled to the target grid.  The technique is very efficient with the additional trick that the down-sampling step can be carried out by the inverse FFT by cutting and reshaping the Fourier grid.

The algorithm of \citet{JascheAliasing} is limited only by the memory needed to apply the FFT on the high resolution grid, so it is competitive with common position-space cell assignment schemes.
However, the sharp cut in k-space leads to an extended, oscillatory distribution in position-space.  Such a decentralised and unphysical cell assignment is undesirable for estimation of the density field.  

An alternative approach was taken by \citet{Cui08} who introduce the use of Daubechies wavelets to approximate the ideal filter while keeping compactness in position-space.  We do not consider this technique here because the centre of mass of the wavelets adopted is offset from the particle position and so they are not well suited for estimation of the density field.

As a compromise, we test a smooth cutoff with a k-space filter with the form:
\begin{equation}\label{eq:window}
W(k_x,k_y,k_z) = \prod_{i = (x,y,z)}  \frac{1}{1+|k_i/k_N|^\alpha}
\end{equation}
The parameter $\alpha$ adjusts the sharpness of the cut: we test $\alpha=100$ ({\it Sup hard}) and $\alpha=8$  ({\it Sup soft}).

Fig. \ref{fig:kern} shows the comparison of the mass assignment schemes including the super-sampling technique.  We see that the hard cut (Sup hard) leads to 
ringing in position-space (the transform is the sinc function).  The amplitude is significantly reduced when the parameter $\alpha=8$ (Sup soft) and the first peak is more compact than CIC.  
In k-space, the filter remains nearly 1 to $\sim 0.7 k_N$ while NGP, CIC and TSC functions are dropping.  Beyond the Nyquist frequency, the soft cut-off kernel drops faster than TSC without
oscillatory behaviour.  The bottom panel of  \ref{fig:kern} shows the cumulative power $S(R) = \int_0^R |W(r)|^2 dr$.  We see that soft-cut off scheme has compactness similar to CIC and NGP, while the latter schemes 
severely damp the total power.

Fig. \ref{fig:alias} compares the aliased power in the power spectrum measurement.  The top four panels show the shot noise power multiplied by the kernels.  The principal contribution ($n=0$) and the first harmonic ($n=-1$) are plotted along with the full sum of all harmonics.  We see for instance, that the power measured with the NGP scheme mixes power from all harmonics.  The hard cut-off filter performs nearly ideally with no aliasing effect.
The soft cut-off also performs well matching the aliasing characteristics of the TSC scheme. 

To implement the soft cut-off filter we use the following practical super-sampling recipe:
\begin{enumerate}
\item Assign particles to a grid with resolution increased by a factor $f$.  The assignment is done with the CIC scheme.  We set $f=8$.
\item Transform the field to Fourier space with the FFT.
\item Multiply Fourier modes by the k-space filter.
\item Transform back to position space with the inverse-FFT.
\item Take a sample of the grid every $f$ cells to down-sample to the target resolution.
\end{enumerate}

\section{Gibbs sampler}\label{appendix:method}
Here we give the algorithms used to sample from the conditional probability functions.

\subsection{Sampling the density field}\label{appendix:WF}

Using a galaxy survey we count galaxies in a given sample $l$ and construct the spatial field $\bN_l$ which is a vector of length $n_{cells}$.
We will write the set of $m$ galaxy samples (\eg different luminosity and redshift bins) as
$\{\bN_l\} \equiv \{\bN^{l_1},\bN^{l_2},...,\bN^{l_m}\}$. Each sample $l$ has a corresponding mean density $\Nbar_l$ and bias $b_l$.

We write the conditional probability for the underlying density field $\bdelta$ as
\begin{equation}
\begin{split}
p(\bdelta | \{\bN_l\},\{\bar{N}_l\},\{b_l\},\Sig) &\propto p(\{\bN_l\} | \bdelta, \{\bar{N}_l\},\{b_l\},\Sig)p(\bdelta|\Sig) \\ 
& =  \prod_{l=1}^{N_{samples}} p(\bN_l | \bdelta, \bar{N}_l,b_l,\Sig)  p(\bdelta|\Sig) \label{eq:condind}
\end{split}
\end{equation}
To write the last line we use the property that the number counts of different samples $l$ are conditionally independent but depend  on the common underlying density field.

For clarity we now explicitly index the cells with subscript $i$.  We  adopt a Gaussian model for the number counts and write the log of the likelihood  $\log p \propto \chi^2$ as
\begin{equation}\label{eq:likelihood}
\begin{split}
-2\log p(N_l | \Nbar_l,b_l,\delta) =  \\
\sum_{i=1}^{n_{cells}}  \frac{ \left(   N_{l,i} - \Nbar_l  w_{l,i} \left( 1 + b_l D(z_i) \delta_i\right)    \right)^2 }{\sigma^2_{l,i}} +  \log (2\pi  \sigma^2_{l,i}) .
\end{split}
\end{equation}
The variance of counts is generalised as $\sigma^2_{l,i}$ and may be different from the Poisson expectation $\bar{N}w$ due to the mask and anti-aliasing filter.  However, we neglect noise correlations between cells which would introduce off-diagonal terms.  The sums are carried out over cells with variance $\sigma^2_{l,i} > 0$.

Next we consider the density field prior.  We set a Gaussian model giving
\begin{equation}
-2\log p(\bdelta|\Sig) = \sum_{j=1}^{n_{cells}} \sum_{i=1}^{n_{cells}} \left( \Sinv \right)_{ij} \delta_i \delta_j  + \log{\left( 2\pi \det{\Sig} \right)} \label{eq:densprior}
\end{equation}
The correlation function of $\delta$ is given by the anisotropic covariance matrix $\Sig$.  In  Fourier space the matrix is diagonal and given by Eq. \ref{eq:pk}.  When writing the posterior we now drop the terms that do not depend on $\delta$ giving
\begin{equation}
\begin{split}
-2\log p(\bdelta | \{\bN_l\},\{\bar{N}_l\},\{b_l\},\Sig) = \\
				\sum_{l=1}^{m_{samples}}\sum_{i=1}^{n_{cells}}  \frac{\left(    N_{l,i} - \Nbar_l  w_{l,i} \left( 1 + b_l D_i \delta_i\right)        \right)^2 }{\sigma^2_{l,i}}  
                                   + \sum_{j=1}^{n_{cells}}  \left( \Sinv \right)_{ij} \delta_i \delta_j   \label{eq:posterior}
\end{split}
\end{equation}
Differentiating the log posterior with respect to $\bdelta$ we find the equation for the maximum a posteriori estimator which is also called the Wiener filter.  The estimate
$\hat{\bdelta}$ is given by
\begin{equation}\label{eq:wf}
\begin{split}
\sum_{l=1}^{m_{samples}}  \frac{\left( \Nbar_l w_{l,i} b_l D_i \right)^2}{\sigma^2_{l,i}} \hat{\bdelta}_i & + \sum_{j=1}^{n_{cells}}  \left(\Sinv\right)_{i,j} \hat{\delta}_j    =\\
&    \sum_{l=1}^{m_{samples}} \frac{\Nbar_l w_{l,i} b_l D_i}{\sigma^2_{l,i}} \left(N_{l,i} - \Nbar_l w_{l,i}\right)
\end{split}
\end{equation}

It is informative to point out how the Wiener filter operation combines the galaxy subsamples.  The right hand side of Eq. \ref{eq:wf} shows the weighted combination, given by
\begin{equation}
\sum_{l=1}^{m_{samples}}  \frac{\Nbar_l w_{l,i} b_l D_i}{\sigma^2_{l,i}} \left(N_{l,i} - \Nbar_l w_{l,i}\right)
\end{equation}
where we consider cell $i$ and the sum is over galaxy samples indexed by $l$.
The subsamples are being weighted by their relative biases $b_l$.  This form of  weighting for the density field  was derived by \citet{Cai11} and is optimal in the case of Poisson sampling but also matches the weights for the power spectrum \citep{Percival04}.

To generate a residual field $\delta_r$ with the correct covariance we follow the method of \citet{Jewell04}.  We draw two sets of Gaussian distributed random variables with zero mean and unit variance: $w_{1,i}$ and $w_{2,i}$.  The residual field is then found by solving the following linear equations.
\begin{equation}
\zeta_i = \sum_{l=1}^{m_{samples}}  \frac{ \Nbar_l w_{l,i} b_l D_i }{\sigma_{l,i}}  w_{1,i}
\end{equation}
\begin{equation}
\eta_i = \sum_{j=1}^{n_{cells}}  \left(\sqrt \Sinv \right)_{i,j} w_{2,j}
\end{equation}
\begin{equation}\label{eq:cr}
\sum_{l=1}^{m_{samples}}  \frac{\left( \Nbar_l w_{l,i} b_l D_i \right)^2}{\sigma^2_{l,i}} \bdelta_{r,i}  + \sum_{j=1}^{n_{cells}}  \left(\Sinv\right)_{i,j} \delta_{r,j}    = \zeta_i + \eta_i
\end{equation}
The final constrained realisation is given by the sum $\delta_{CR}=\hat{\delta}+\delta_r$.

The terms including the signal covariance matrix are computed in Fourier space where the matrix is diagonal.  The transformation is done using the Fast fourier transform algorithm.  We solve Eq. \ref{eq:wf} and Eq. \ref{eq:cr} using the linear conjugate gradient solver {\tt bicgstab} provided in the SciPy python library (\url{http://www.scipy.org/}).

\subsection{Sampling the signal}\label{appendix:signal}
The posterior distribution for the signal is 
\begin{equation}
p(S | \delta) \propto p(\delta | S)p(S)
\end{equation}
We take a flat prior leaving $p(S | \delta) \propto p(\delta | S)$ which is given by Eq. \ref{eq:densprior}.  This is recognised
as an inverse-Gamma distribution for $S$ which can be sampled directly \citep{Kitaura08,Jasche10-SDSSBayes}.

We parametrise the signal in Fourier space as the product of the real-space power spectrum and redshift-space distortion model (Eq. \ref{eq:pk}).
We sample the parameters in two steps.  First we fix $\beta$ and $\sigma_v$ and draw a real-space power spectrum from the inverse-gamma distribution.

The normalisation of the power spectrum is fixed by 
\begin{equation}
\sigma_R^2 = \int d^3 k P(k) |W(k)|^2
\end{equation}

In the regime where shot noise is more important than cosmic variance, this method of sampling is inefficient.  An alternative sampling scheme was proposed to improve the convergence \citep{Jewell09,Jasche10-SDSSBayes}.  The new power spectrum is drawn making a step in both $P$ and $\delta$ such that $\delta^{s+1} = \sqrt{P^{s+1}/P^s} \delta^s$.  The consequence is that the conditional posterior $p(P|\delta)$ is unchanged but the data likelihood is modified through $\delta$.  We perform a sequence of Metropolis-Hastings steps to jointly draw $P$ and $\delta$.  We alternate between the two modes for sampling $P$ on each step of the  Gibbs sampler.

The redshift-space distortion parameters $\beta$ and $\sigma_v$ are next sampled jointly.  These parameters are strongly degenerate on the scales we consider.  To efficiently sample these we
compute the eigen decomposition of the Hessian matrix to rotate into a coordinate system with two orthogonal parameters.  We evaluate the joint probability over a two-dimensional grid in the orthogonal space.
Finally we use a rejection sampling algorithm to jointly draw values of $\beta$ and $\sigma_v$.

\subsection{Sampling galaxy bias}\label{appendix:bias}
The bias is sampled in the manner described by \citet{Jasche13}.  The conditional probability distribution for the bias of a given galaxy sample is a Gaussian with mean given by:
\begin{equation}
\mu_b = \frac{\sum \bar{N}_i w_i D_i \delta_i (N_i-\bar{N}_i)/\sigma^2_i} {\sum_i \left(\bar{N}_iw_i D_i \delta_i\right)^2/\sigma_i^2}
\end{equation}
and variance
\begin{equation}
\sigma^2_b = \frac{1} {\sum_i \left(\bar{N}_iw_i D_i \delta_i\right)^2/\sigma_i^2}
\end{equation}
After drawing a value from a Gaussian distribution with the given mean and variance, it is limited to the range $0.5<b<4.0$.

\subsection{Sampling mean density}\label{appendix:meandensity}
We take a Poisson model for the mean density.  Taking a uniform prior we have  
\begin{eqnarray}
p(\bar{N} | \delta,N,b,S) &\propto&  p(N | \delta, b, \bar{N},S)p(\bar{N})\\
&\propto& \prod_i \frac{\left[\Nbar w_i \left(1+b D \delta\right)\right]^{N_i}}{N_i!}  e^{-\Nbar w_i \left(1+b D_i \delta_i \right)}
\end{eqnarray}
Keeping only terms that depend on $\bar{N}$ we have
\begin{equation}
\log p(\bar{N}) = \sum_i N_i \log \bar{N} - \Nbar w_i \left(1+b D \delta_i \right)
\end{equation}
We draw a sample from this distribution by computing the likelihood over a grid of values and using a rejection sampling algorithm.

\section{Convergence analysis}\label{appendix:conv}
\begin{figure}
\includegraphics{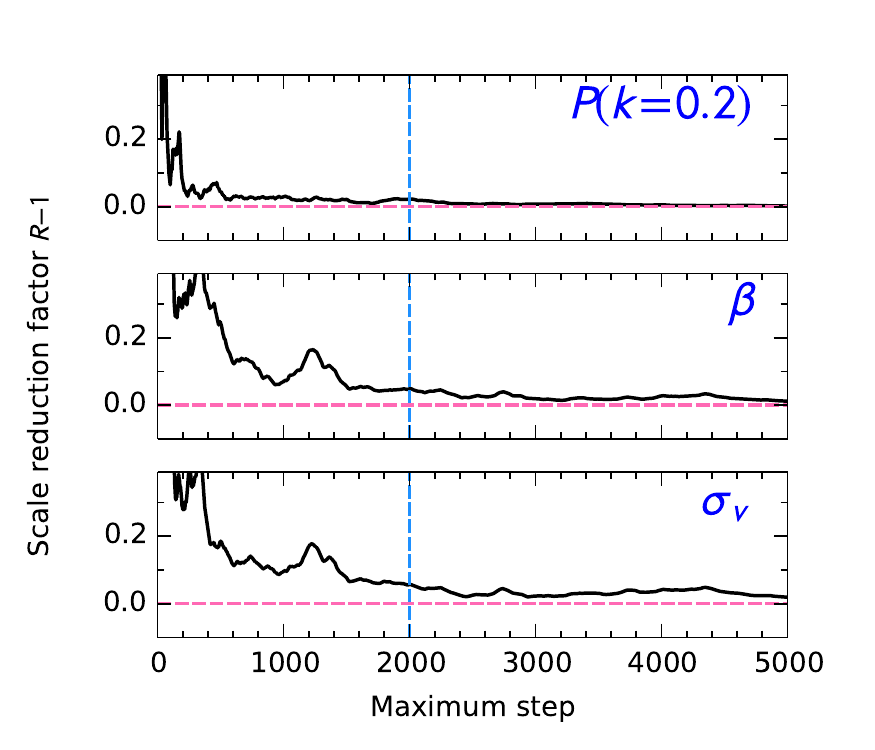}
\caption{The Gelman-Rubin convergence diagnostic $R$ computed from a single mock catalogue with 10 chains.   \label{fig:conv}}
\end{figure}

The Gelman-Rubin  convergence diagnostic \citep{Gelman92,Brooks98} involves running multiple independent Markov chains and comparing the single-chain variance to the between-chain variance.  The scale reduction factor $R$ is the ratio of the two variance estimates.  In Fig. \ref{fig:conv} we show this convergence diagnostic for three parameters: the power spectrum at $k=0.2$\mpch, the distortion parameter $\beta$ and the velocity dispersion $\sigma_v$.  These parameters are the slowest to converge.  The $R$ factor is computed up to a given maximum step $i$ in the chain after discarding the first $i/2$ steps.  We consider the chain to be properly `burned-in' when $R<1.2$.  Based on this diagnostic we set the maximum step in our analysis to be 2000 and discard the first 1000 steps

\end{document}